\documentclass[prl,aps,amssymb,showpacs,twocolumn,superscriptaddress,times,10pt]{revtex4-1}
\usepackage{amsmath}
\usepackage{amssymb}
\usepackage{amsfonts}
\usepackage{listings}
\usepackage{enumerate}
\usepackage{latexsym}
\usepackage{bm}
\usepackage{graphicx}
\usepackage{braket}
\usepackage[pdftex,colorlinks=false]{hyperref}
\usepackage{xcolor}
\usepackage{verbatim}

\usepackage{times}

\newcounter{defcounter}
\newcommand{\bs}[1]{\ensuremath{\boldsymbol{#1}}}

\renewcommand{\dag}{\dagger}
\newcommand{\nodag}{{\vphantom\dag}}

\newcommand\varpm{\mathbin{\vcenter{\hbox{%
  \oalign{\hfil$\scriptstyle+$\hfil\cr
          \noalign{\kern-.3ex}
          $\scriptscriptstyle({-})$\cr}%
}}}}
\newcommand\varmp{\mathbin{\vcenter{\hbox{%
  \oalign{$\scriptstyle({+})$\cr
          \noalign{\kern-.3ex}
          \hfil$\scriptscriptstyle-$\hfil\cr}%
}}}}

\begin{document}

\title{Topology and Magnetism in the Kondo Insulator Phase Diagram}

\author{Michael~Klett}
\address{
Institute for Theoretical Physics and Astrophysics, University of W\"urzburg, Am Hubland, D-97074 W\"urzburg, Germany
}

\author{Seulgi~Ok}
\address{
Department of Physics, University of Zurich, Winterthurerstrasse 190, 8057 Zurich, Switzerland
}

\author{David~Riegler}
\address{
Institute for Theoretical Physics and Astrophysics, University of W\"urzburg, Am Hubland, D-97074 W\"urzburg, Germany
}

\author{Peter W\"olfle}
\address{
Institute for Theory of Condensed Matter and Institute for Nanotechnology, Karlsruhe Institute of Technology, 76021 Karlsruhe, Germany
}
	
\author{Ronny~Thomale}
\email{rthomale@physik.uni-wuerzburg.de}
\address{
Institute for Theoretical Physics and Astrophysics, University of W\"urzburg, Am Hubland, D-97074 W\"urzburg, Germany
}
	
\author{Titus~Neupert}
\email{titus.neupert@uzh.ch}
\address{
Department of Physics, University of Zurich, Winterthurerstrasse 190, 8057 Zurich, Switzerland
}

\begin{abstract}

Topological Kondo insulators are a rare example of an interaction-enabled topological phase of matter in three-dimensional crystals -- making them an intriguing but also hard case for theoretical studies.
Here, we aim to advance their theoretical understanding by solving the paradigmatic two-band model for topological Kondo-insulators using a fully spin-rotation invariant slave-boson treatment. Within a mean-field approximation, we map out the magnetic phase diagram and characterize both antiferromagnetic and paramagnetic phases by their topological properties. Among others, we identify an antiferromagnetic insulator that shows, for suitable crystal terminations, topologically protected hinge modes. 
Furthermore, Gaussian fluctuations of the slave boson fields around their mean-field value are included in order to establish the stability of the mean-field solution through computation of the full dynamical susceptibility.  	
		
\end{abstract}

\maketitle

\textit{Introduction} --- 
Landau's theory of spontaneous symmetry breaking and 
topological phenomena are often cited as two antipodal concepts by which phases of matter can be organized. 
However, in strongly correlated topological systems, which are surprisingly rare in three-dimensional systems, they can show an intriguing interplay. 
One of the paradigmatic examples are topological Kondo insulators~\cite{dzero2010topological,dzero2016topological}, in which $d$ and $f$ electron band partially overlap and hybridize. 
This band overlap sources the band-inversion central to topological band theory, while correlations from the strongly localized $f$ electrons induce a robust hybridization gap between these bands and thus bring about an insulating ground state.

Several materials fall in this category, with SmB$_6$ being the best-studied example. 
The nature of its ground state is still under debate despite intense experimental investigations, with evidence for a topological~\cite{nikolic2014, alexandrov2015, roy2015, Nakajima2016, thomson2016, kwchang2017} as well as for a non-topological scenario~\cite{yzhou2017}, while some works even report indications for a metallic state~\cite{berman1983effect, beille1983suppression, gabani2004insulator}.
Another point of controversy is the magnetic order of SmB$_6$, with indications for paramagnetic (PM)~\cite{biswas2014low}, anti-ferromagnetic~\cite{yzhou2017} and surface-ferromagnetic phases. The presence of magnetism would crucially influence the topological classification of the material.
Jointly, these results demonstrate that SmB$_6$ and more broadly topological Kondo insulators are at a nontrivial intersection of topology, symmetry breaking, and correlated phenomena.

This renders the systems not only a highly relevant but also intrinsically hard case for theoretical studies.
Analytical and numerical methods are challenged by a strongly correlated interplay between localized and delocalized as well as spin and orbital degrees of freedom. In this work, we adopt
 Kotliar-Ruckenstein's formulation of the slave-boson formalism \cite{kotliar_new_1986}, which matches with the Gutzwiller approximation in infinite dimensions, to map out its magnetic and topological phase diagram.
We use Kotliar and Ruckenstein's scheme among the many variants of slave-boson formulations, because it has been extended to a spin-rotation invariant description~\cite{woelfle_spin_rotation_1989,woelfle_spin_1992}, including Gaussian fluctuations~\cite{Li1991,woelfle_spin_1997,PAM_Wuerzburg}, and refer to it shortly as slave-bosons from here on.

In the slave-boson treatment, one replaces the fermionic creation and annihilation operators by pseudo-fermionic ones applied by an auxiliary bosonic field.
The bosonic fields are chosen such that the fermionic interaction terms are replaced by quadratic bosonic terms in the action, while the resulting pseudo-fermionic fields also only enter quadratically.
When \emph{local} constraints are imposed through Lagrange multiplier fields to constrain the values of bosonic fields to  physical ones, one obtains an exact representation of the original problem.
The calculation is then facilitated by the approximation to impose these constraints only on average, instead of on every individual site. We introduce this mean-field ansatz for the bosonic fields such that it can yield both magnetic and PM solutions. Following Ref.~\cite{PAM_Wuerzburg} one can acquire the expressions with the physical information of the original system, e.g., effective mass, spin, and charge susceptibility.
Together with the periodic Anderson model, Kondo systems have been considered as a particularly suitable target to prove the accuracy of slave-bosons for their physics including interacting electrons as well as hybridized orbitals.
	
In this work, we numerically implement the analytical representations from the slave-boson formalism, and present the phase diagram for a model that resembles the low-energy physics in SmB$_6$, but can be seen as paradigmatic for generic three-dimensional Kondo models with cubic and time-reversal symmetry.
We find a total of seven phases, that are magnetically or topologically distinct, by tuning the strength of the electron-electron interactions as well as the on-site energy of the $f$-electrons. Besides various PM insulating topological phases, we find two topologically distinct insulating phases with $(\pi,\pi,\pi)$ antiferromagnetic order. 

	\textit{Model and method} \label{sec2} --- 
We start with a short exposition of our model and the main ingredients of the spin-rotation invariant slave-boson formalism.
The Hamiltonian  introduced in \cite{legner2014topological} as a minimal model for SmB$_6$ reads
\begin{subequations}
\begingroup	
\allowdisplaybreaks
	\begin{align}
	H 
	=&
	H_0 +
	H_{\text{hyb}} +
	H_{\text{int}}, \label{eq:1hbtevvv}
	\end{align}
\endgroup
with
\begingroup	
\allowdisplaybreaks
	\begin{align}
	\nonumber
	H_0 
	=&
	\sum_{i} \widetilde{\bs{f}}_i^{~\dagger} \epsilon_{f} \widetilde{\bs{f}}_i^\nodag 
	-\sum_{\langle i,j \rangle }\left( \widetilde{\bs{f}}_i^{~\dagger} t^f \widetilde{\bs{f}}_j^\nodag + \bs c^\dagger_i t^d \bs c^\nodag_j + \text{H.c.} \right)
	\\ 
	-&\sum_{\langle \langle i,j \rangle \rangle}\left( \widetilde{\bs{f}}_i^{~\dagger} t^{f'} \widetilde{\bs{f}}_j^\nodag + \bs c^\dagger_i t^{d'} \bs c^\nodag_j + \text{H.c.} \right) 
	- \sum_{i}  \mu_{0} \tilde{n}_i^\nodag,
	\\ 
	H_{\text{hyb}}
	=&
	\sum_\alpha \sum_{\langle i,j \rangle_\alpha } \textrm{i} V\left( \widetilde{\bs{f}}_i^{~\dagger} \tau^\alpha \bs c^\nodag_j + \bs c^\dagger_i \tau^\alpha \widetilde{\bs{f}}_j^\nodag + \text{H.c.} \right),
	\\ 
	H_{\text{int}} =& ~U \sum_{i} \tilde{f}_{i,\uparrow}^{\dagger} \tilde{f}_{i,\uparrow} \tilde{f}_{i,\downarrow}^{\dagger} \tilde{f}_{i,\downarrow}
	\label{eq:2hbtevvv},
	\end{align}
\endgroup
\label{eq:1}
\end{subequations}
representing $f$- and $d$-electron hopping on a simple cubic lattice, their hybridization and the local repulsive Hubbard interaction, respectively.	
	The spinors $\widetilde{\bs f}_{i}^{~\dagger} := (\tilde{f}_{i,\uparrow}^{\dagger},\tilde{f}_{i,\downarrow}^{\dagger})$ 
	and $\bs c_i^\dagger = (c_{i,\uparrow}^{\dagger},c_{i,\downarrow}^{\dagger})$
	are formed by  the creation operators of $d$- ($f$-) electrons $c_{i,\sigma}^{\dagger}$ ($\tilde{f}_{i,\sigma}^{\dagger}$) with spin $\sigma \in \{ \uparrow, \downarrow \}$
	and $\tilde{n}_i^\nodag = \widetilde{\bs f}_{i}^{~\dagger} \widetilde{\bs f}_{i}^{\nodag} + \bs c_{i}^{\dagger} \bs c_{i}^{\nodag} $ represents the local density of all electrons at site $i$. 
	We denote by $\langle i,j \rangle_\alpha$ a nearest-neighbor bond in $\alpha$-direction and $\tau^\alpha$ as the $\alpha$-th component of the vector of Pauli matrices $\boldsymbol{\tau} := (\tau^1,\tau^2,\tau^3)$ acting in spin space. 
	The notations for nearest ($\langle \cdot \rangle$) and next-to-nearest ($\langle \langle \cdot \rangle \rangle$) neighbor pairs of sites are adopted in conventional form.
	
	We consider a half-filled band structure throughout and choose $t^{d}=1$, $t^{f}=-0.1$, $t^{d'}=-0.4$, $t^{f'}=0.04$, and $V=0.5$, all energy scales will be given in units of $t^{d}$.
The form of Eq.~\eqref{eq:1} and the parameters chosen account for negligible interactions as well as bigger hopping amplitudes among the $d$-electrons as compared to the $f$ electrons.
We chose the relative energy shift between $f$ and $d$ orbitals $\epsilon_{f}$ and the interaction strength $U$ as free parameters as a function of which we will map out the phase diagram. In Ref.~\cite{legner2014topological} the phase diagram has been constructed from a simpler slave-boson mean field solution that is not spin-rotation invariant and does not account for magnetic phases.
We find several gapped non-magnetic phases, except at the lines of topological phase transitions as well as gapped and metallic phases in the $(\pi,\pi,\pi)$ antiferromagnetically ordered regime. The magnetic instability occurs in close proximity to the location of the topological transitions identified in  Ref.~\cite{legner2014topological}.

	\textit{Slave-boson representation}\label{Sec:Method} ---
	To account for interactions, we apply the slave-boson representation of the
	operators  $\widetilde{\bs f}_{i}^{~(\dagger)}$, originally introduced by Kotliar and Ruckenstein \cite{kotliar_new_1986}
	which has been generalized to be spin-rotation invariant (SRI)~\cite{woelfle_spin_rotation_1989,woelfle_spin_1992} and
	to consider fluctuations around the a PM saddle point \cite{Li1991,woelfle_spin_1997}. We introduce the bosonic 
	fields $e_{i},d_{i},p_{i,0}$ and $\boldsymbol{p}_{i} := (p_{i,1},p_{i,2},p_{i,3})$, labeling empty, doubly and singly occupied states, respectively, i.e.,
	\begin{equation}
	\begin{split}
	e_{i}^{\dagger} |\textrm{vac} \rangle := |\textrm{vac} \rangle \ , \quad
	d_{i}^{\dagger} f_{i,\uparrow}^{\dagger} f_{i,\downarrow}^{\dagger} |\textrm{vac} \rangle := \tilde{f}_{i,\uparrow}^{\dagger} \tilde{f}_{i,\downarrow}^{\dagger} |\textrm{vac} \rangle \ , \\
	\frac{1}{2} 
	\sum_{\tilde{\alpha}=0}^3 \sum_{\sigma'} 
	(p_{i,\tilde{\alpha}}^{\dagger} \tau^{\tilde{\alpha}})_{\sigma \sigma'}
	f_{i,\sigma'}^{\dagger}|\textrm{vac} \rangle := \tilde{f}_{i,\sigma}^{\dagger} |\textrm{vac} \rangle \ ,\\
	\end{split}
	\label{eq:4cnerstb}
	\end{equation}
	with $|\textrm{vac} \rangle$ being the vacuum state and the operators $\bs f_{i}^{\dagger} := (f_{i,\uparrow}^{\dagger},f_{i,\downarrow}^{\dagger})$ 
	being a new set of auxiliary fermionic operators. The unity matrix is denoted by $\tau^0$.
	The Hilbert space defined by the slave particle operators has to be projected on to the physical Hilbert space by the application of constraints (see Supplementary Material ~SM),
	which are inserted via Lagrange multiplier terms in the Lagrangian by introducing five new fields $i\alpha_{i}$, $i\beta_{i,0}$, and $i\beta_{i,\alpha}$ per site $i$.\\
	\textit{Mean field approximation}\label{sec2.B} ---
	We apply a mean field ansatz, incorporating a static spin spiral with ordering vector $\boldsymbol Q$ of a possible magnetic ground state. Following Ref.~\cite{PhysRevB.48.10320}, we replace the 
	bosonic fields at the lattice site labeled by $i$ with lattice vector $\bs r_i$ by $ b_i \rightarrow  b$, where $b_i$ represents any of the fields $e_i,d_i,p_{0,i},\alpha_i,\beta_{0,i}$,
\begin{align}
\label{mfansatz}
\begin{split}
\bs{p}_i \rightarrow 
\begin{pmatrix}
\cos(\phi_i)\\
\sin(\phi_i)\\
0
\end{pmatrix}
p, \quad \text{and} \quad
i \bs \beta_i \rightarrow 
\begin{pmatrix}
\cos(\phi_i)\\
\sin(\phi_i)\\
0 
\end{pmatrix}
\beta.
\end{split}
\end{align}
Here we have $\phi_i = \bs Q \cdot \bs r_i$,  $\beta \in \mathbb{R}$ and  $b,p \in \mathbb{R}_0^+$. 
	Within this mean field ansatz the free energy per site is given by (see SM)
\begin{equation}\label{freeenergy}
\begin{split}
\frac{F}{N}= &-\frac TN\sum_{\nu = 1}^8\sum_{\boldsymbol k} \ln\left[1+e^{-\epsilon_{\boldsymbol k,\boldsymbol Q}^\nu/T}\right]+U d^2\\&+n\mu_0-\beta_0(2d^2+ p^2 + p_0^2)-2\beta p_0 p ,
\end{split}
\end{equation}
	where $\epsilon_{\boldsymbol k,\boldsymbol Q}^\nu$ are the renormalized eigenvalues of the effective mean field Hamiltonian, that implicitly depend on the slave boson fields, and $T$ is the temperature. 
	The index $\nu$ is a combined label for the spin, orbital and sublattice degrees of freedom. The filling is fixed at $n=2$. 
As shown in SM, the free energy~Eq.~\eqref{freeenergy} is invariant under global SO(3) rotations of all $\bs p_i$.
	
	We distinguish mean field solutions with $p=0$, describing a PM state, and $p\neq 0$, signaling magnetic order. They are obtained by minimizing the 
	free energy with respect to $e,d,p_0$ and $p$ while maximizing with respect to $\beta, \beta_0$ and $\mu_0$.
Since there is plethora of possible ordering vectors $\bs Q$ to consider, we first calculate the PM mean field by explicitly setting $p=\beta=0$ and then perform a stability analysis of the saddle point by expanding the action $S$ up to second order in fluctuations of the bosonic fields.

\begin{figure*}
\includegraphics[width=1\textwidth]{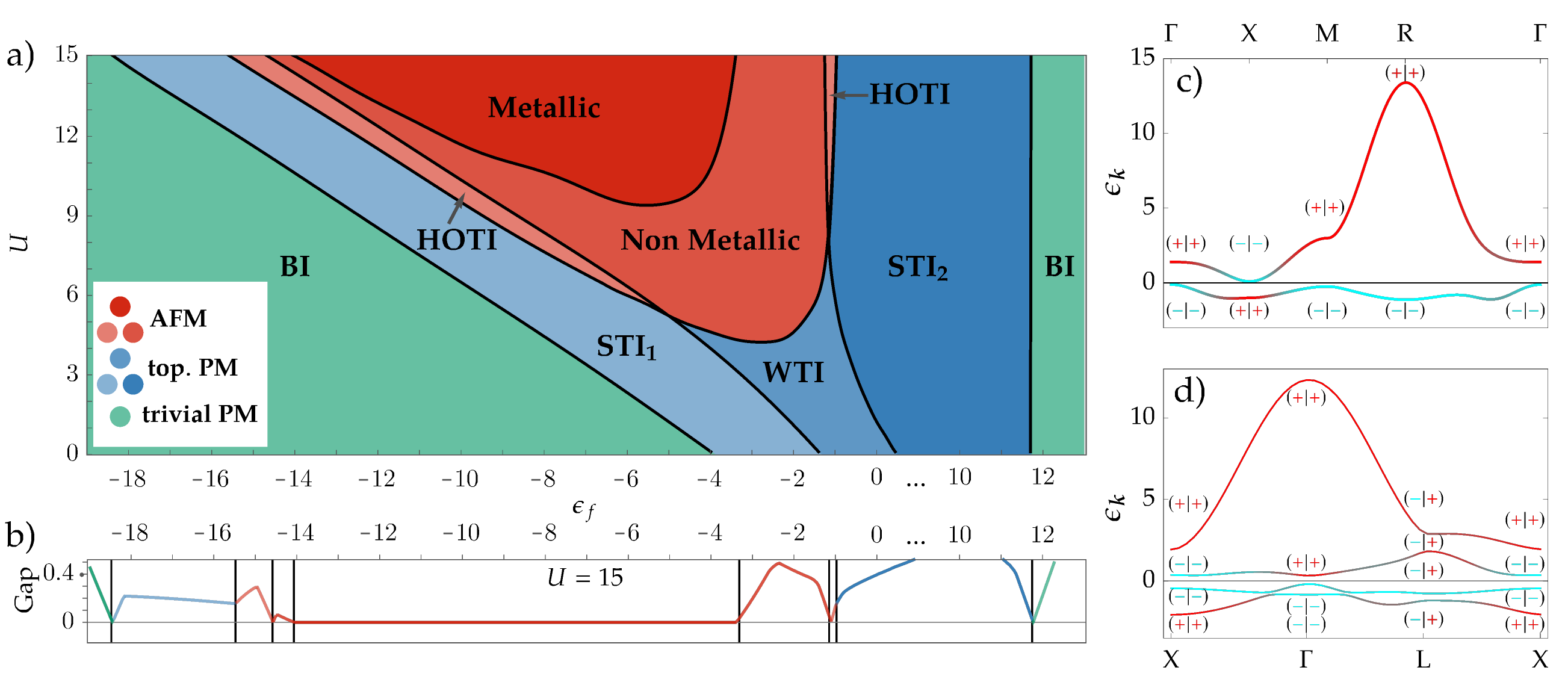}
		\caption{ a):
			Topological phase diagram in the parameter space of $(\epsilon_f,U)$.
			BI, WTI, STI$_1$, and STI$_2$ phases are found in PM region, whereas a non metallic, metallic, and HOTI phases appear in AFM region.
			b): The gap plot is for the path of $U=15$, where one can see the topological gap closings as well as the metallic AFM region.
			c): Bulk band structure of SmB$_6$ for $\epsilon_f=-12$ and $U=10$ (STI$_1$) on the high symmetry path $\Gamma-\mbox{X}-\mbox{M}-\Gamma$ in the first BZ of a simple cubic lattice. Cyan (red) colored lines represent strong $f$ ($c$)
			occupation and $(+|-)$ the inversion eigenvalues of the Kramer's pair.
			d): Bulk band structure of SmB$_6$ for $\epsilon_f=-12$ and $U=12$ (HOTI) on the high symmetry path $\mbox{X}-\Gamma-\mbox{L}-\mbox{X}$ in the first BZ of a face centered cubic lattice. Cyan (red) colored lines represent strong $f$ ($c$)
			occupation and $(+|-)$ the inversion eigenvalues of the Kramer's pair. Red band weights in the occupied lower part of the spectrum of c) and d) indicate band inversion.
		}
		\label{fig:1fvajrfee}
\end{figure*}

	\textit{Fluctuations around the paramagnetic saddle point} \label{sec2.C} ---
	Within this expansion, the charge and spin susceptibility in momentum space are obtained as
	\begin{equation}
	\begin{split}
	\chi_{\textrm{c}}(q) &:= 
	\frac{1}{2} \langle \delta n(-q)\, \delta n(q) \rangle \ , \\
	\chi_{\textrm{s}}^{\alpha\beta} (q) &:= 
	\frac{1}{2} \langle \delta S^{\alpha}(-q)\, \delta S^{\beta}(q) \rangle,
	\end{split}
	\end{equation}
	where $S^\alpha = p_0p_\alpha$ represents the spin density operator projected to direction $\alpha$ and $n = 2d^2 + p_0^2 + \bs p^2$ is the 
	charge density operator expressed in bosonic fields. 
Through an analytical calculation for the mode at hand, we find $\chi_{\textrm{s}}^{\alpha\beta} (q)$ to be proportional to the unit matrix and will hence refer to it as a scalar function $\chi_{\textrm{s}} (q)$~\cite{PAM_Wuerzburg}. 
Here, $q:= (\bs q,\omega)$ is the four-vector of the wave vector $\bs q$ and 
	frequency $\omega$, and $\langle \cdot \rangle$ is the thermal expectation value of Gaussian fluctuations around the saddle point.
We adopt the expressions for the susceptibilities from \cite{Hubbard_Wuerzburg,PAM_Wuerzburg}. A sign change in the real part in the zero frequency limit signals the onset of spontaneous charge or magnetic order. In the parameter domain explored we do not find any charge instabilities.
However, the real part of $\chi_{\textrm{s}}$ exhibits a sign change in the two-dimensional parameter space $(\epsilon_{f},U)$, implying magnetic order. 
We compare the critical $U$ at vanishing $\omega$ for the instability with $\bs q =(\pi,\pi,\pi)$ to that of other ordering vectors.
None of the latter led to an instability at smaller $U$, thus we conclude that the magnetic ordering vector is $\bs Q =(\pi,\pi,\pi)$~(see SM).
The magnetic phase boundary can be seen in Fig.~\ref{fig:1fvajrfee}. 
Investigating possible (incommensurate) magnetic or charge instabilities which could emerge on the AFM band structure deeper in the phase would require a susceptibility analysis of the magnetic band structure, which is beyond the scope of this paper.	 
In the supplementary material we show the general form of a mean field with arbitrary ordering vector $\bs Q$.

\textit{Topology of PM and antiferromagnetic band structures} \label{sec2.D} ---
The solutions of the mean field equations for $\boldsymbol{Q}=(\pi,\pi,\pi)$ yield magnetic and non-magnetic domains in the parameter space of $(\epsilon_{f},U)$. 
The resulting PM phase boundary coincides with the one obtained by analyzing the fluctuations around the paramagnetic saddle point, separating the phase diagram in a PM (blue/green) and AFM (red) domain in Fig.~\ref{fig:1fvajrfee}. The remaining phase boundaries indicate either a change of topology or a metal to insulator transition.
	
Using the renormalized band structure, we can study the band topology. 
In the PM case, the effective hopping Hamiltonian (see SM) features antiunitary time-reversal $\mathcal{T}: \big( \mathcal{T} c^{(\dagger)}_{\sigma,\boldsymbol{k}} \mathcal{T}^{-1}, \mathcal{T} f^{(\dagger)}_{\sigma,\boldsymbol{k}} \mathcal{T}^{-1} \big) \mapsto \big( \sigma c^{(\dagger)}_{-\sigma,-\boldsymbol{k}},\sigma f^{(\dagger)}_{-\sigma,-\boldsymbol{k}} \big)$ and unitary inversion $\mathcal{I}: \big( \mathcal{I} c^{(\dagger)}_{\sigma,\boldsymbol{k}} \mathcal{I}^{-1}, \mathcal{I} f^{(\dagger)}_{\sigma,\boldsymbol{k}} \mathcal{I}^{-1} \big) \mapsto \big( c^{(\dagger)}_{\sigma,-\boldsymbol{k}}, -f^{(\dagger)}_{\sigma,-\boldsymbol{k}} \big)$ symmetry, where $c^{(\dagger)}_{\sigma,-\boldsymbol{k}}$, represents any fermionic operator with quantum numbers $(\sigma,-\boldsymbol{k})$.
Because $(\mathcal{T}\mathcal{I})^2=-1$, bands are doubly degenerate at every $\boldsymbol{k}$.
We can define the $\mathbb{Z}_2$-valued
strong ($\nu^{\textrm{PM}}_{0}$) and weak ($\nu^{\textrm{PM}}_{\alpha}$) topological indices~\cite{fu2007topological, fu2007topological2, legner2014topological, qi2008topological}:
\begin{subequations}
\begingroup	
\allowdisplaybreaks
\begin{align}
\text{strong:} \quad \prod_{j=1}^{8} \prod_{n \in \textrm{occupied}} \xi[n,\Gamma_{j}] &= (-1)^{\nu^{\textrm{PM}}_{0}} \ , \\
\text{weak:}
\prod_{j;k_{\alpha}=\pi} \prod_{n \in \textrm{occupied}} \xi[n,\Gamma_{j}] &= (-1)^{\nu^{\textrm{PM}}_{\alpha}} \ .
\label{eq:8rvcxvbgl}
\end{align}
\endgroup
\end{subequations}
	Here, $\Gamma_{j}$ represents the time-reversal-invariant momenta (TRIM), which are 
	$\Gamma^{\textrm{PM}} = (0,0,0)$, $\textrm{X}^{\textrm{PM}} \in \{ (\pi,0,0), (0,\pi,0), (0,0,\pi) \}$, $\textrm{M}^{\textrm{PM}} \in \{ (\pi,\pi,0), (0,\pi,\pi), (\pi,0,\pi) \}$ 
	and $\textrm{R}^{\textrm{PM}} = (\pi,\pi,\pi)$ in the first Brillouin zone in the simple cubic lattice. $\xi[n,\Gamma_{j}]$ is the inversion eigenvalue of the $n$th Kramer's 
	pair at the $\Gamma_{j}$ point. Due to the cubic symmetry in the PM phase, all the weak indices are equivalent, i.e., $\nu^{\textrm{PM}}_{x} = \nu^{\textrm{PM}}_{y} = \nu^{\textrm{PM}}_{z}$.
	Therefore, we can maximally have four topologically distinct phases given by $(\nu^{\textrm{PM}}_{0},\nu^{\textrm{PM}}_{\alpha}) \in \{ (0,0) , (0,1) , (1,0) , (1,1) \}$.
	We denote these phases by BI, WTI, STI$_{1}$, and STI$_{2}$, respectively.	
We illustrate the factors of the strong index $\nu^{\textrm{PM}}_{0}$ for the STI$_{1}$ phase at $(\epsilon_f,U)=(-12,10)$ in Fig.~\ref{fig:1fvajrfee}c): the eigenvalues $\xi[1,\Gamma_j]$ are $-1$ ($+1$)
for the blue (red) colored portions of the energy bands, such that $\xi[1,\Gamma^{\textrm{PM}}]=\xi[1,\textrm{M}^{\textrm{PM}}]=\xi[1,\textrm{R}^{\textrm{PM}}]=-1$, but $\xi[1,\textrm{X}^{\textrm{PM}}]=1$ and consequently $\nu^{\textrm{PM}}_{0}=1$.
	
In the magnetically ordered phases with ordering vector $\bs Q=(\pi,\pi,\pi)$, the real space primitive unit cell is doubled in size, since each site $i$ is surrounded by six neighboring sites with opposite spin expectation value.
In this anti-ferromagnetic phase, the system has a pseudo-time-reversal symmetry $\mathcal{T}'$ as the combination of $\mathcal{T}$ and the translation by $\hat{\boldsymbol{r}}_\alpha$ to a nearest neighbor in $\alpha$-direction. 
Moreover, inversion symmetry $\mathcal{I}$ is retained. 
We find four doubly degenerate bands.
The change in the unit cell from simple cubic to face-centered cubic results in eight new TRIMs, which are 
$\Gamma:=0$, $\textrm{L}:=\boldsymbol{b}_{1}/2$, $\textrm{X}:=(\boldsymbol{b}_{1}+\boldsymbol{b}_{2})/2$.
Of these $\textrm{L}$ and $\textrm{X}$ have, respectively, four and three partners obtained by $C_{4}$ rotations.
Here, $\boldsymbol{b}_{1}:=(-\pi,\pi,\pi)$, $\boldsymbol{b}_{2}:=(\pi,-\pi,\pi)$, and $\boldsymbol{b}_{3}:=(\pi,\pi,-\pi)$ in units of $|\hat{\boldsymbol{r}}_\alpha|^{-1}$ are the three primitive reciprocal lattice vectors of the fcc structure.
Hence, the strong index
\begin{equation}
\begin{split}
\nu^{\textrm{AFM}}: 
& \prod_{n \in \textrm{occupied}} \xi[n,\Gamma]\, \xi[n,\textrm{X}] 
= (-1)^{\nu^{\textrm{AFM}}}
\end{split}
\end{equation}
corresponding to Eq.~\eqref{eq:8rvcxvbgl} remains $\mathbb{Z}_{2}$-valued. Since there are four $\textrm{L}$ points, the related factor drops out of the product. As shown in Fig.~\ref{fig:1fvajrfee} each Kramer's pair at the $\textrm{L}$ point has two different inversion eigenvalues. Due to the translation operator in $\mathcal{T}'$, which does not commute with the inversion operator at  $\textrm{L}$, we find the relative phase between the eigenvalues of the two pairs to be equivalent to $\exp(2\text{i}\hat{\boldsymbol{r}}_\alpha\textrm{L})=-1$. whereas the weak index
\begin{equation}
\nu^{\textrm{AFM}}_{\textrm{weak}}: \prod_{n \in \textrm{occupied}}  (\xi[n,\textrm{X}])^{2} = 1 = (-1)^{\nu^{\textrm{AFM}}_{\textrm{weak}}}
\end{equation}
is always trivial.
Therefore, apart from the metallic regime where the size of the indirect gap is negative,
we find a region with the trivial topologically index $\nu^{\textrm{AFM}}=0$, which is labeled Non Metallic,
and a higher-order topological insulator (HOTI) $\nu^{\textrm{AFM}}=1$, exhibiting topological states in two dimensions lower than the bulk.
We illustrate the factors of the index $\nu^{\textrm{AFM}}$ for the HOTI phase at $(\epsilon_f,U)=(-12,12)$ in Fig.~\ref{fig:1fvajrfee}d).

\begin{figure}[t!]
		\centering
		\includegraphics[width=0.46\textwidth]{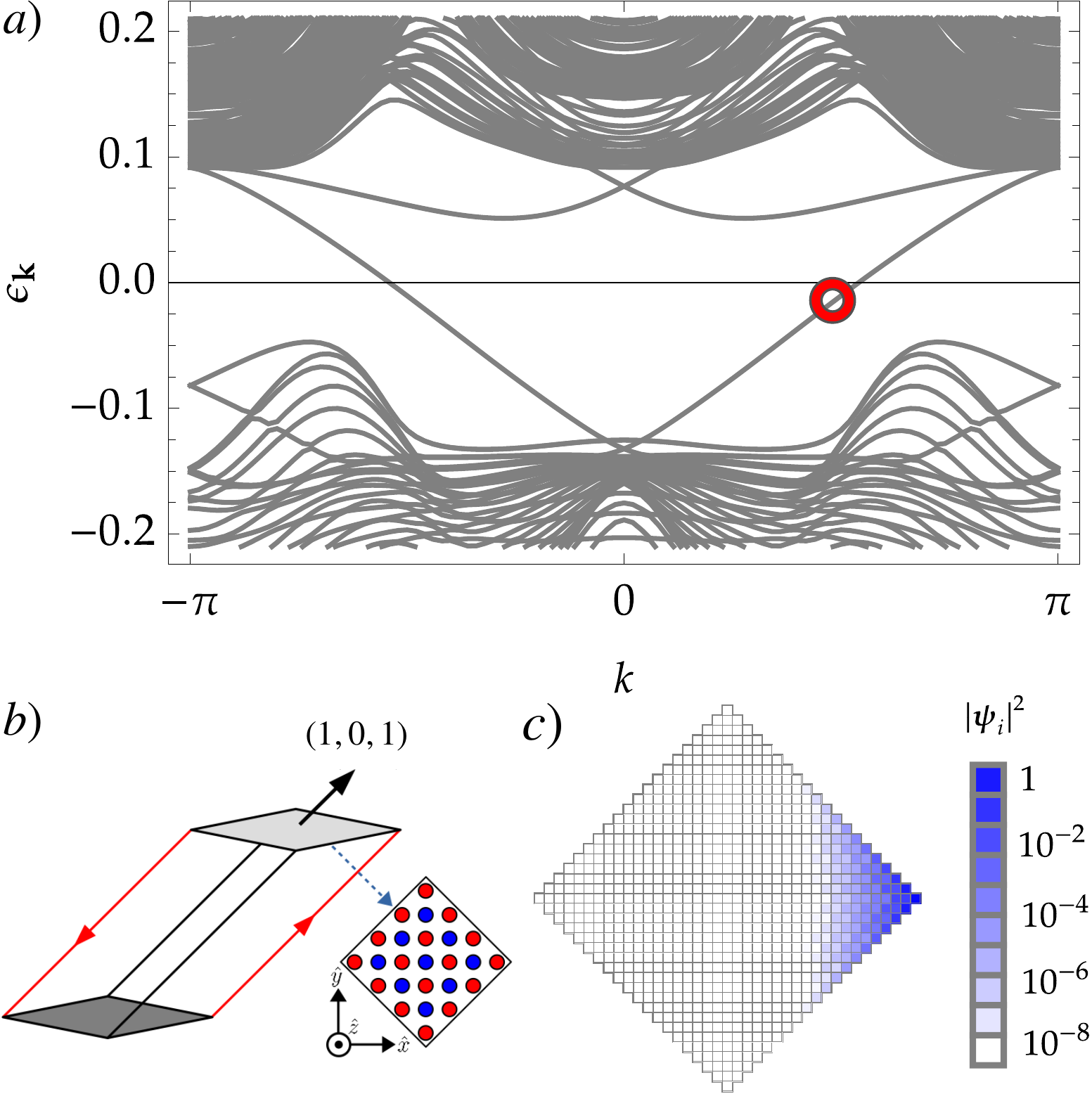}
		\caption{
			a): Band structure in the inclined column geometry depicted in b) at $(\epsilon_f,U)=(-12,12)$ featuring chiral hinge modes.
			b): An example of column geometry where the hinge mode appears.
			The antiferromagnetic spin texture in each plane is shown in the diamond-shaped inset. 
			Red (blue) circles indicate spin $\downarrow$ ($\uparrow$) in the fermionic part of the renormalized Hamiltonian.
			c): Probability distribution in real-space of the hinge state indicated in a) by the red circle.
			$20 \times 20 + 19 \times 19=761$ sites are used for the plot.
		}
		\label{fig:1pvaejcd}
	\end{figure}

The phase with non-trivial topology can further be classified as axion insulator (AXI), which, depending on the orientation of the surfaces, can show gapless chiral hinge modes while the surface and bulk remains gapped. 
These modes are realized in a geometry that preserves $\mathcal{I}$ and breaks $\mathcal{T}'$~\cite{essin2009magnetoelectric, mong2010antiferromagnetic, Schindlereaat0346,ahn2019symmetry, PhysRevLett.119.246401, yue2019symmetry}.
Experimental evidence for such magnetically ordered topological materials was recently observed in MnBi$_2$Te$_4$ and Bi$_2$Se$_3$ thin films~\cite{otrokov2018prediction, gong2019experimental, xu2012hedgehog}. 
The hinge modes in a nanowire geometry are shown in Fig.~\ref{fig:1pvaejcd}. Surfaces that preserve $\mathcal{T}'$ would feature a gapless Dirac cone.

\textit{Phase diagram} ---
To sum up, depending on the parameter set $(\epsilon_f,U)$ we probe, the effective fermionic part of Eq.~\eqref{eq:1} can realize a PM or an antiferromagnetic phase with ordering vector $\boldsymbol{Q}=(\pi,\pi,\pi)$.
There are four topologically distinct sub-phases (BI, WTI, STI$_1$, and STI$_2$) in the PM phase, which are determined by the two topological indices $\nu^{\textrm{PM}_{0}}$ and $\nu^{\textrm{PM}}_{\alpha}$.
On the other hand, there are three sub-phases (metallic, Non Metallic, and HOTI) in the AFM region. We find that the Non Metallic AFM phase features surfaces states, which are not topologically protected (see Fig.~\ref{fig:1fvajrfee}).

We further investigate the possibility of obtaining excitonic states with our formalism (see SM). These charge neutral collective modes have been suggested to account for the long-standing anomalies in SmB$_6$ in several experimental observations and point to the relevance of excitons for the electronic structure of SmB$_6$~\cite{10.1143, knolle2017excitons}. 
However, we do not find any evidence for excitonic states or bands separated from the band continuum within the dynamical spin susceptibility within our slave-boson treatment.
	
\textit{Conclusion} ---
We studied the phase diagram of a paradigmatic model for three-dimensional topological Kondo insulators using the scheme of Kotliar-Ruckenstein's slave-boson representations. To that end, we numerically implemented the analytical expressions of charge and spin susceptibility of a cubic and time-reversal symmetric system.
We obtained a collection of phases in which topological properties and magnetic symmetry breaking are intertwined, yielding, among others, an axion insulator with chiral hinge modes.
Our results provide theoretical guidance to further explore the experimentally observed antiferromagnetism and indications for non-trivial topology in SmB$_6$ and related materials.

\section*{Acknowledgments}
The authors at University of Zurich acknowledge support from the Swiss National Science Foundation (grant number: 200021\_169061) and from the European Union’s Horizon 2020 research and innovation program (ERC-StG-Neupert-757867-PARATOP).
The work in W\"urzburg is funded by the Deutsche Forschungsgemeinschaft (DFG, German Research Foundation) through Project-ID 258499086 - SFB 1170 and through the W\"urzburg-Dresden Cluster of Excellence on Complexity and Topology in Quantum Matter -- \textit{ct.qmat} Project-ID 39085490 - EXC 2147.
The authors also gratefully acknowledge the support of Markus Legner and Jannis Seufert. 

\vspace{5mm}
The authors Michael Klett, Seulgi Ok and David Riegler contributed equally to this work.
\bibliography{Main}
\clearpage

\begin{widetext}
\section*{Supplementary Material}
		
		\subsection{Paramagnetic fluctuation calculations}
		 Fig.~\ref{fig:2varftst} shows the charge- and spin susceptibility yielding from the paramagnetic mean field solution in the $(\epsilon_f,U)$-plane. A detailed derivation of the analytical expressions for the response functions can be found in \cite{PAM_Wuerzburg}.
		 There are no charge instabilities, however, the spin susceptibility features divergences, which occur on the line where $\text{Re} \chi_s$ has a sign change.
		
		This v-shaped spin instability emerges at the lowest interaction $U$ for the ordering vector $\bs Q=(\pi,\pi,\pi)$ which implies antiferromagnetic order in the upper blue colored region where the paramagnetic mean field solution breaks down.
		A more delicate search around $\bs Q=(\pi,\pi,\pi)$ in Fig.~\ref{fig:3ntvsrltytg} and Fig.~\ref{fig:5nwcaweh} gives additional evidences that the magnetic ordering vector is correct.
		\begin{figure}[h]
\centering
\includegraphics[width=0.24\textwidth]{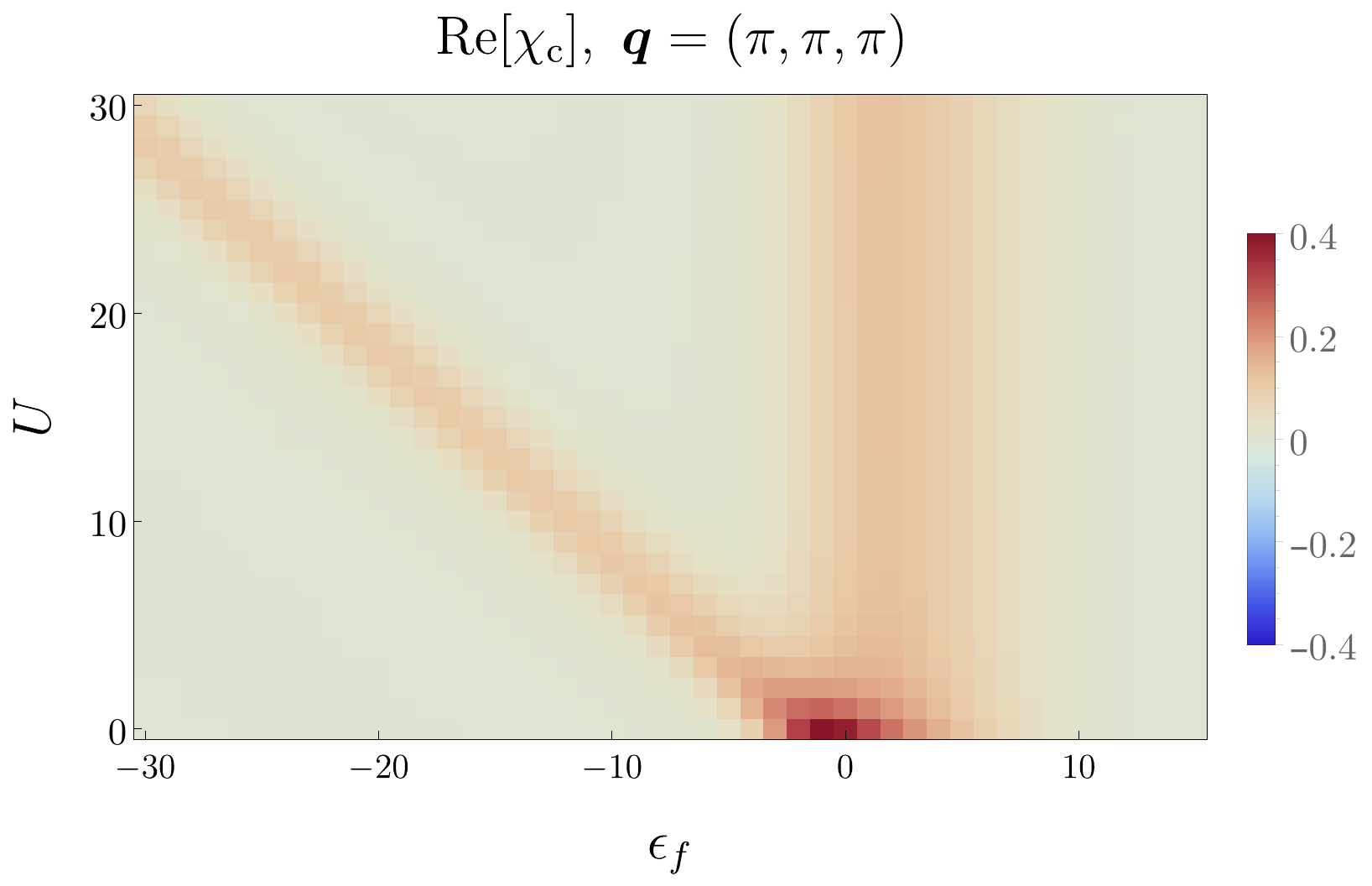}
\includegraphics[width=0.24\textwidth]{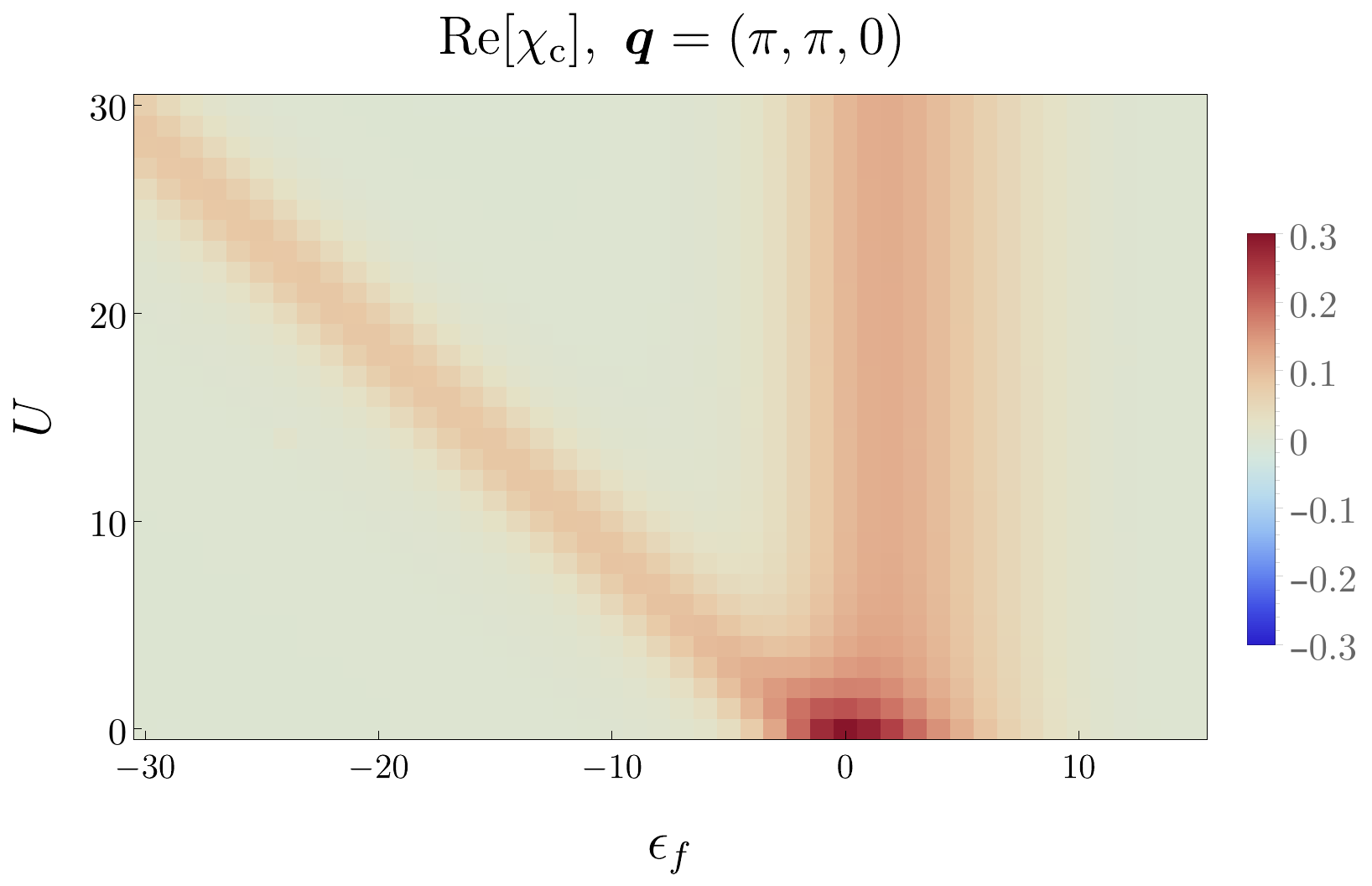}
\includegraphics[width=0.24\textwidth]{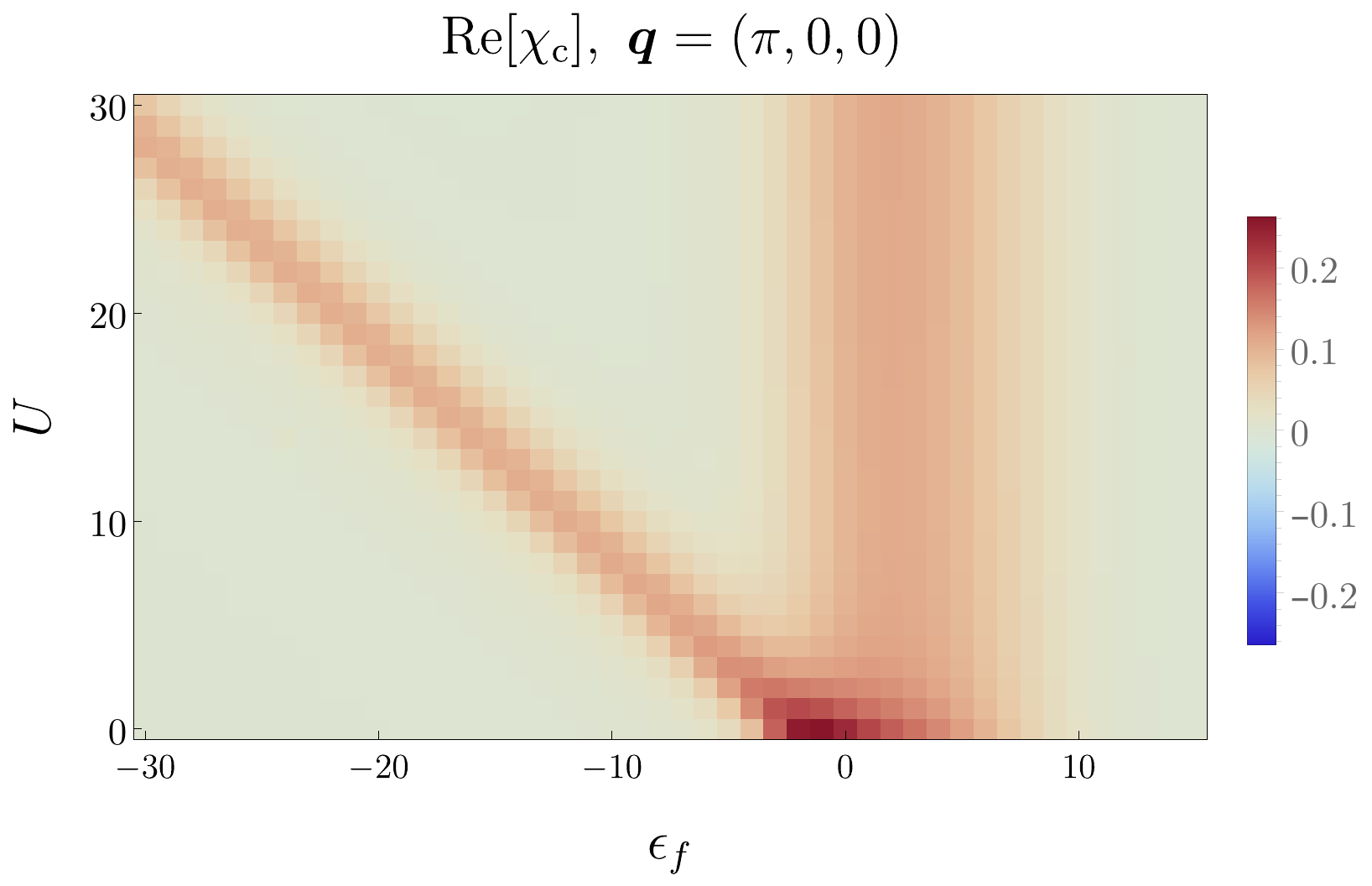}
\includegraphics[width=0.24\textwidth]{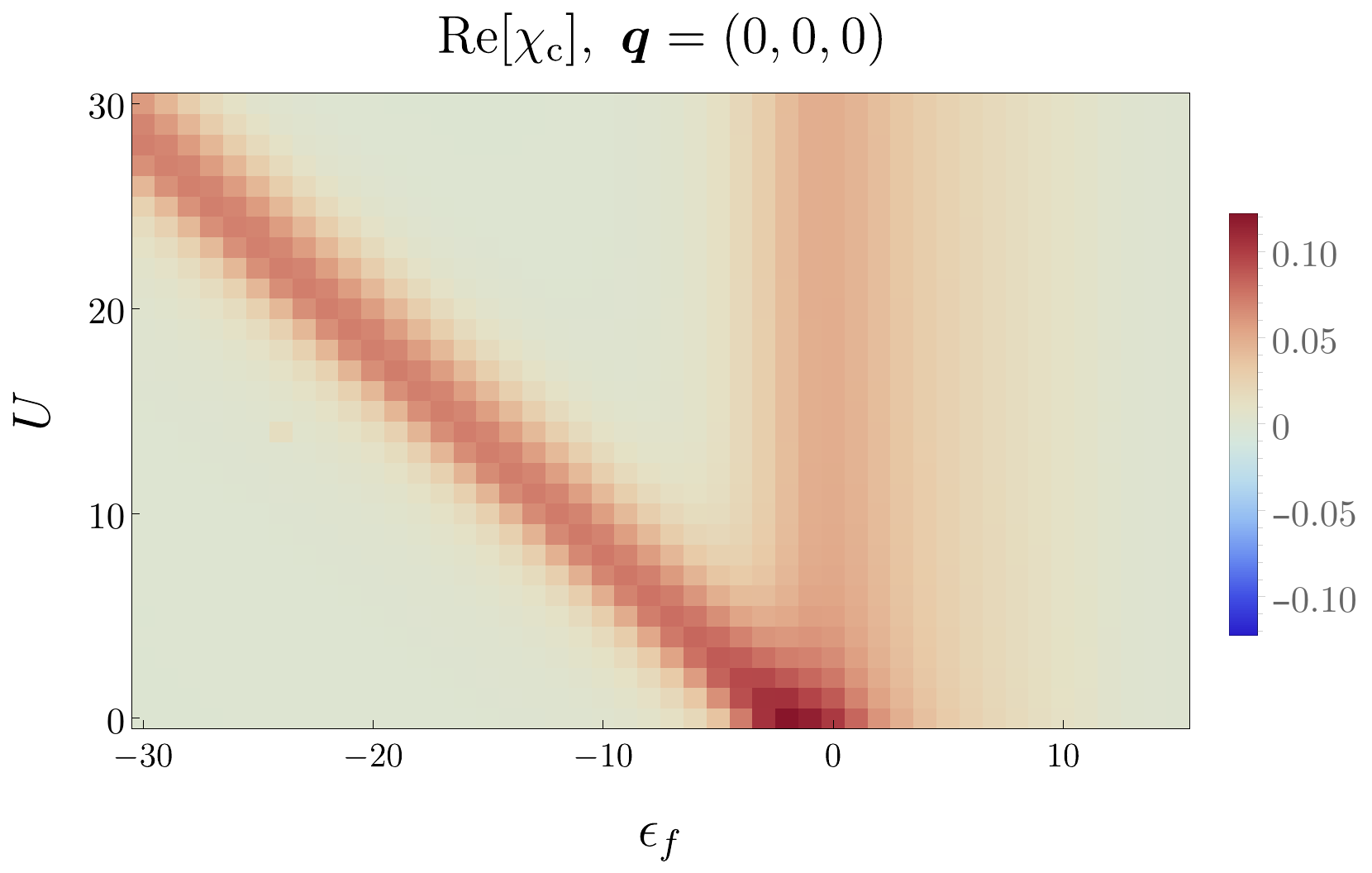} \\
\includegraphics[width=0.24\textwidth]{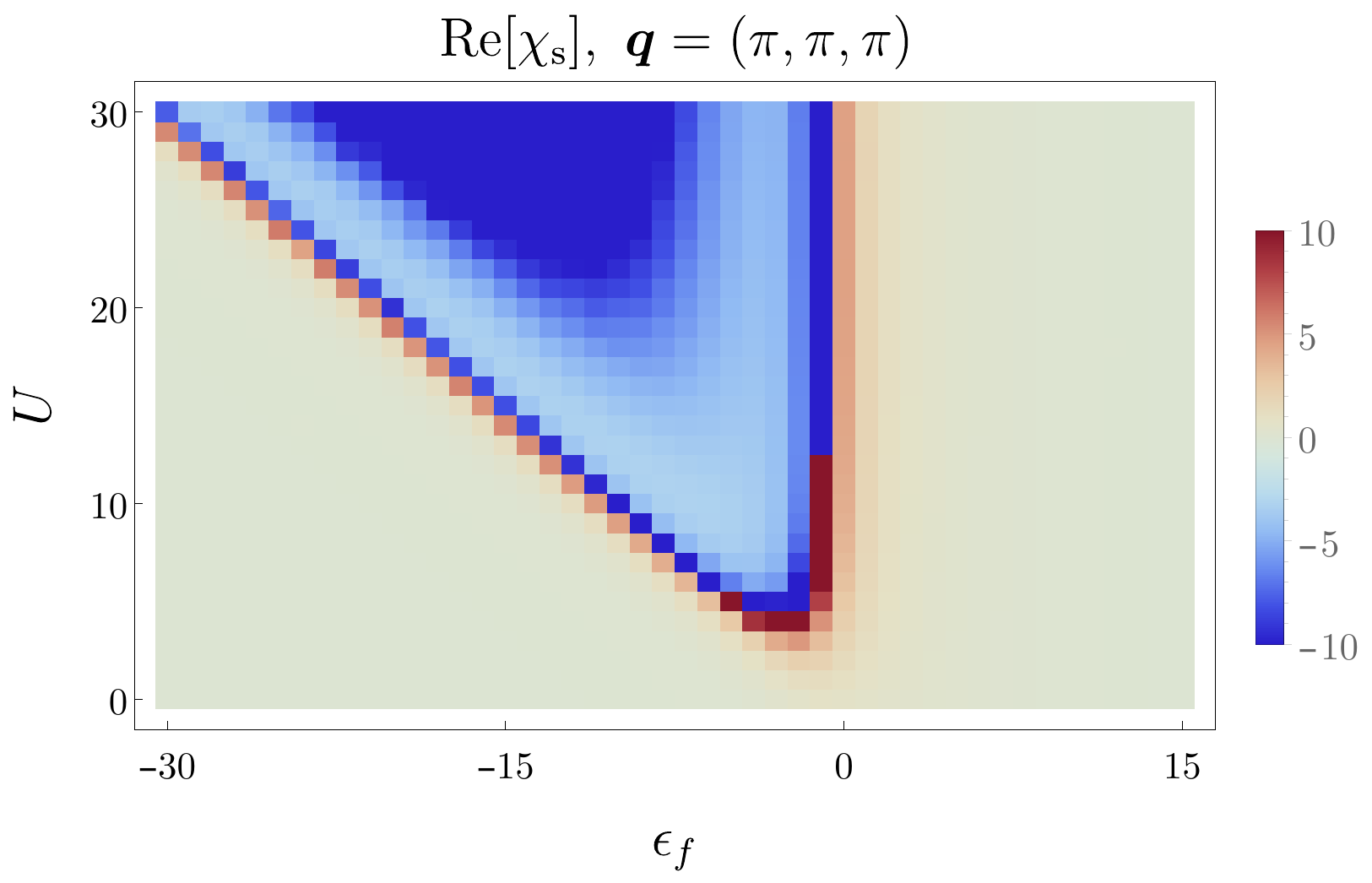}
\includegraphics[width=0.24\textwidth]{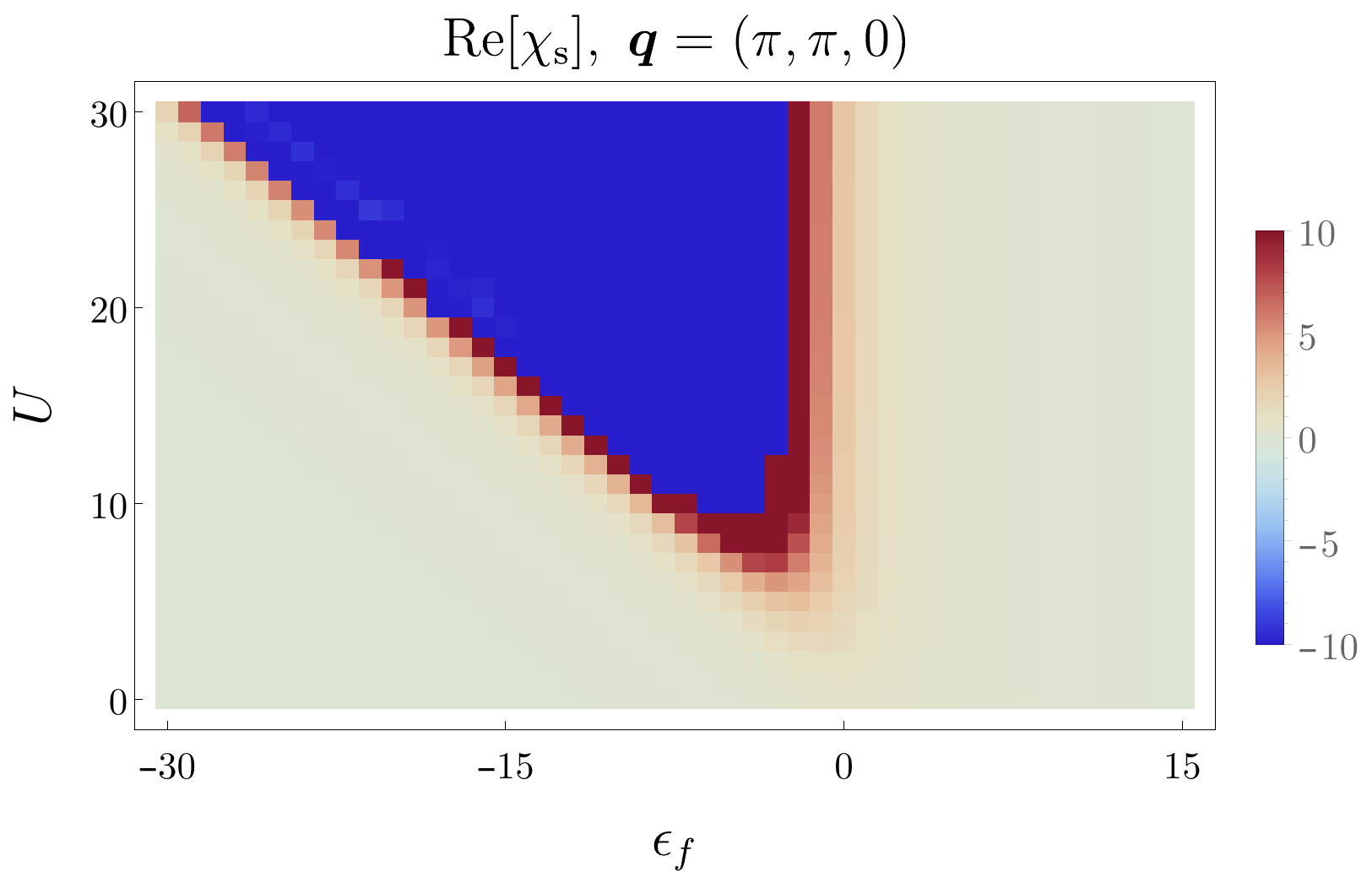}
\includegraphics[width=0.24\textwidth]{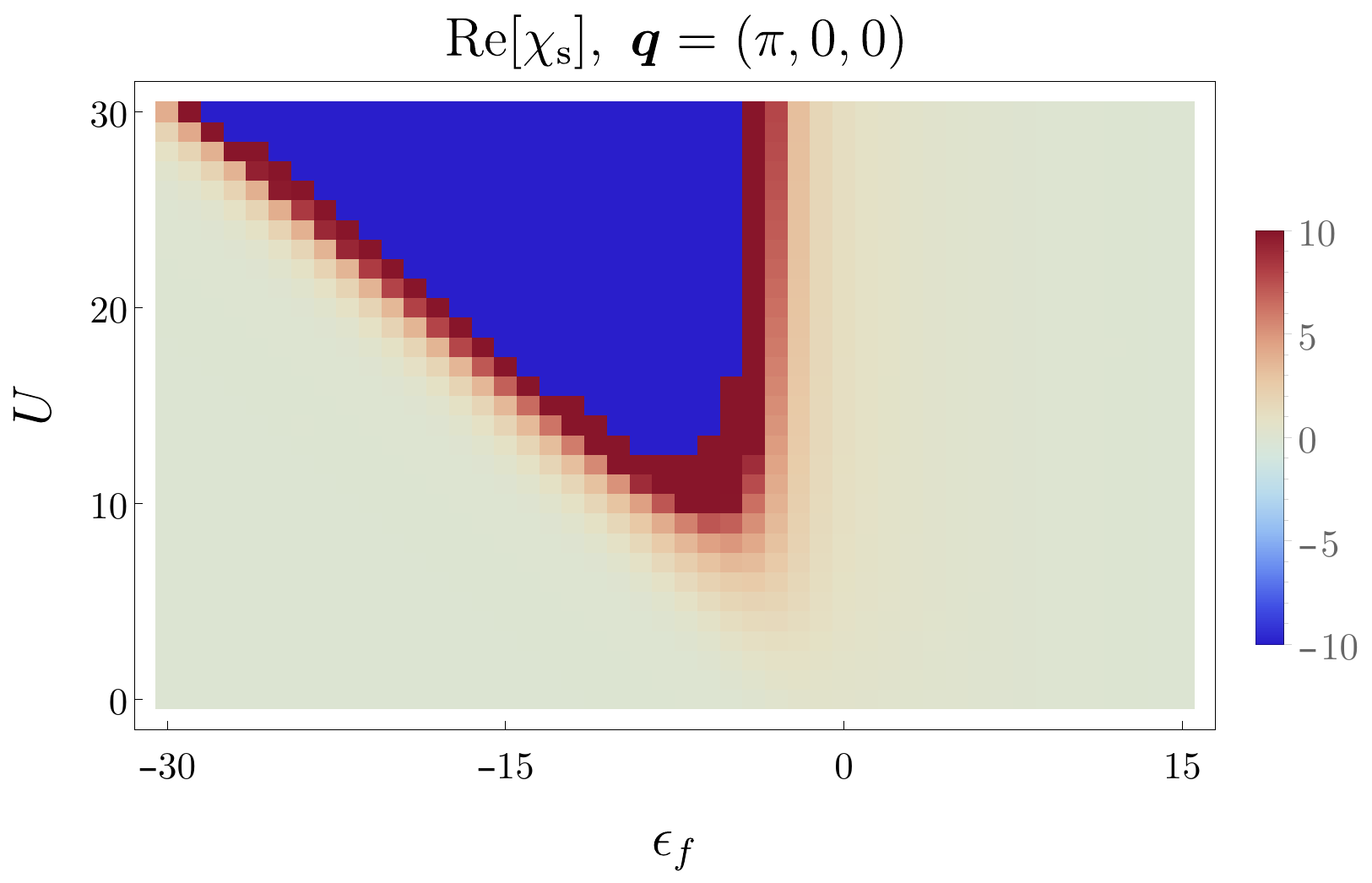}
\includegraphics[width=0.24\textwidth]{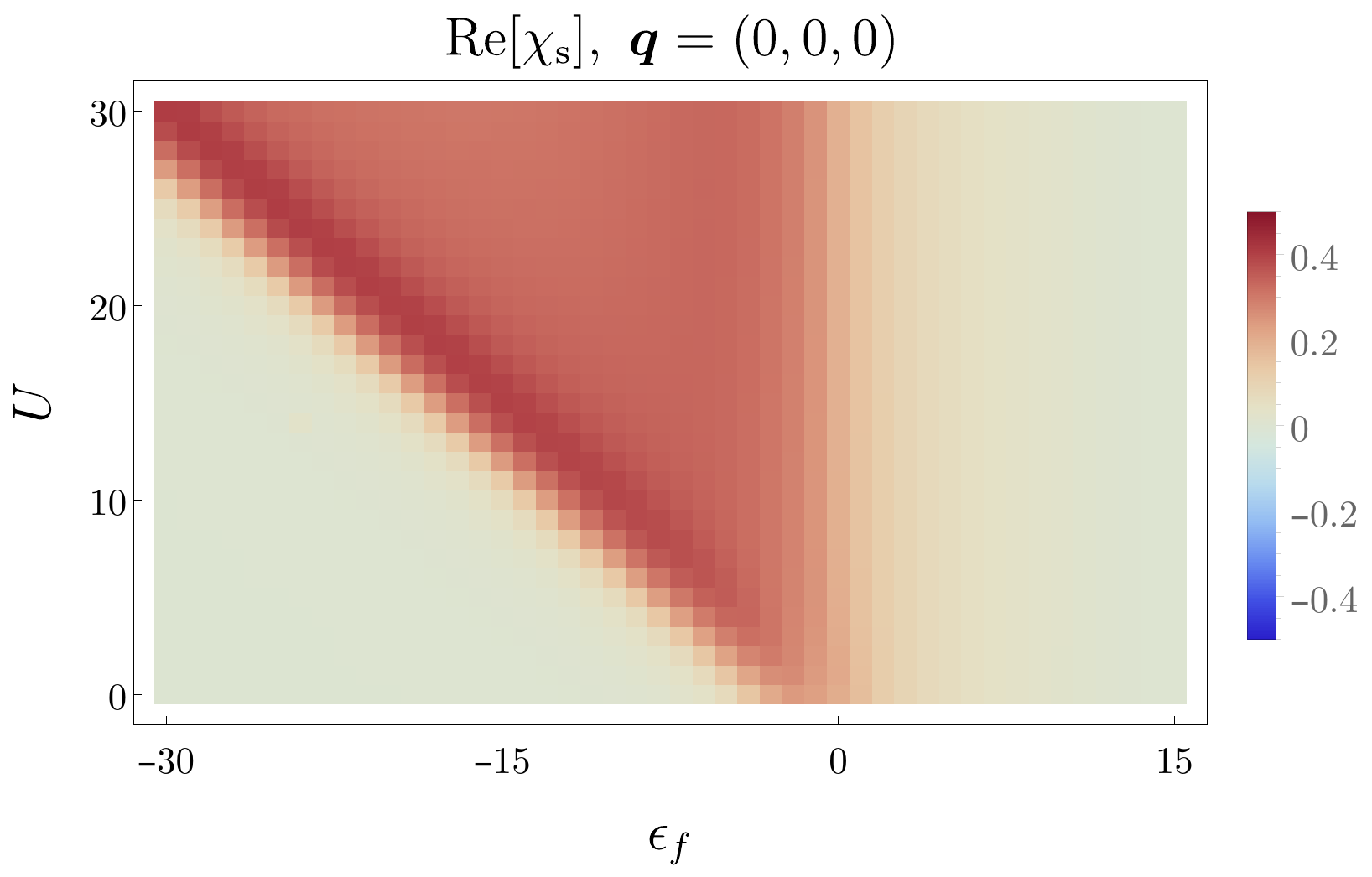}
\caption{
Charge (upper four panels) and spin (lower four panels) susceptibility plots with different ordering vectors $\bs q$.
For the spin susceptibilities, the values bigger than $10$ (smaller than $-10$) are substituted to $10$ ($-10$) for clearer visibility.
The computation has been performed for temperature $T=0.0025$ and Lorenz broadening factor $\eta=0.001$ which yields from the analytical continuation $i\omega \rightarrow \omega + i \eta$ at $\omega=0$.
One finds distinct lines of divergence in $\textrm{Re} [\chi_{s}]$ with $\bs Q=(\pi,\pi,\pi)$, $\bs Q=(\pi,\pi,0)$, and $\bs Q=(\pi,0,0)$.
A sign change in the real part is confirmed on each of the lines.
Since $\bs Q=(\pi,\pi,\pi)$ calls the instability with the lowest value of $U$, we conclude that the system is spontaneously ordered with $\bs Q=(\pi,\pi,\pi)$ in the blue region.
Because the presented susceptibilities are based on the paramagnetic saddle point, they cannot provide a stability analysis of the magnetically ordered region of the phase diagram.
}
\label{fig:2varftst}
\end{figure}

\begin{figure}[h]
\centering
\includegraphics[width=0.24\textwidth]{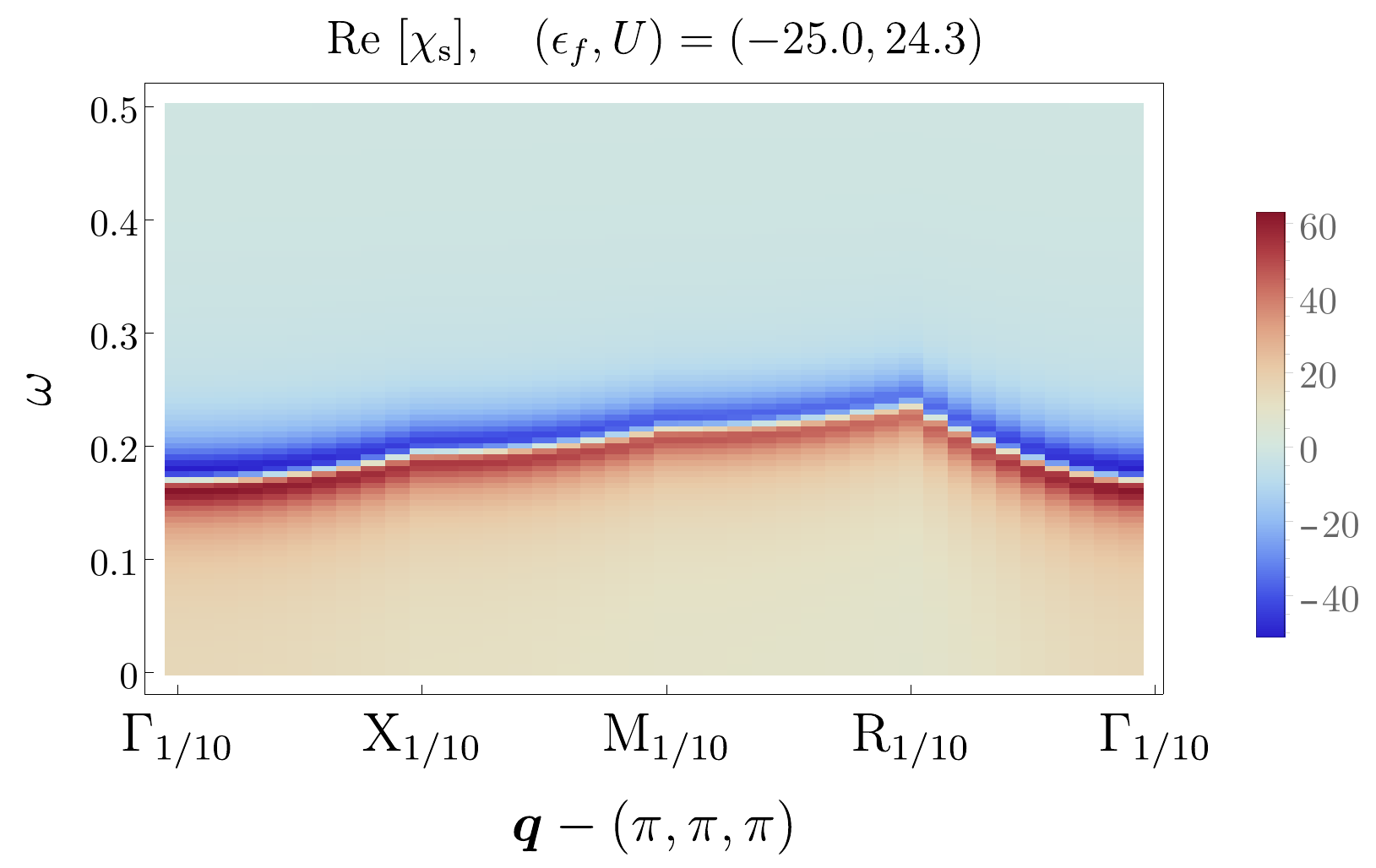}
\includegraphics[width=0.24\textwidth]{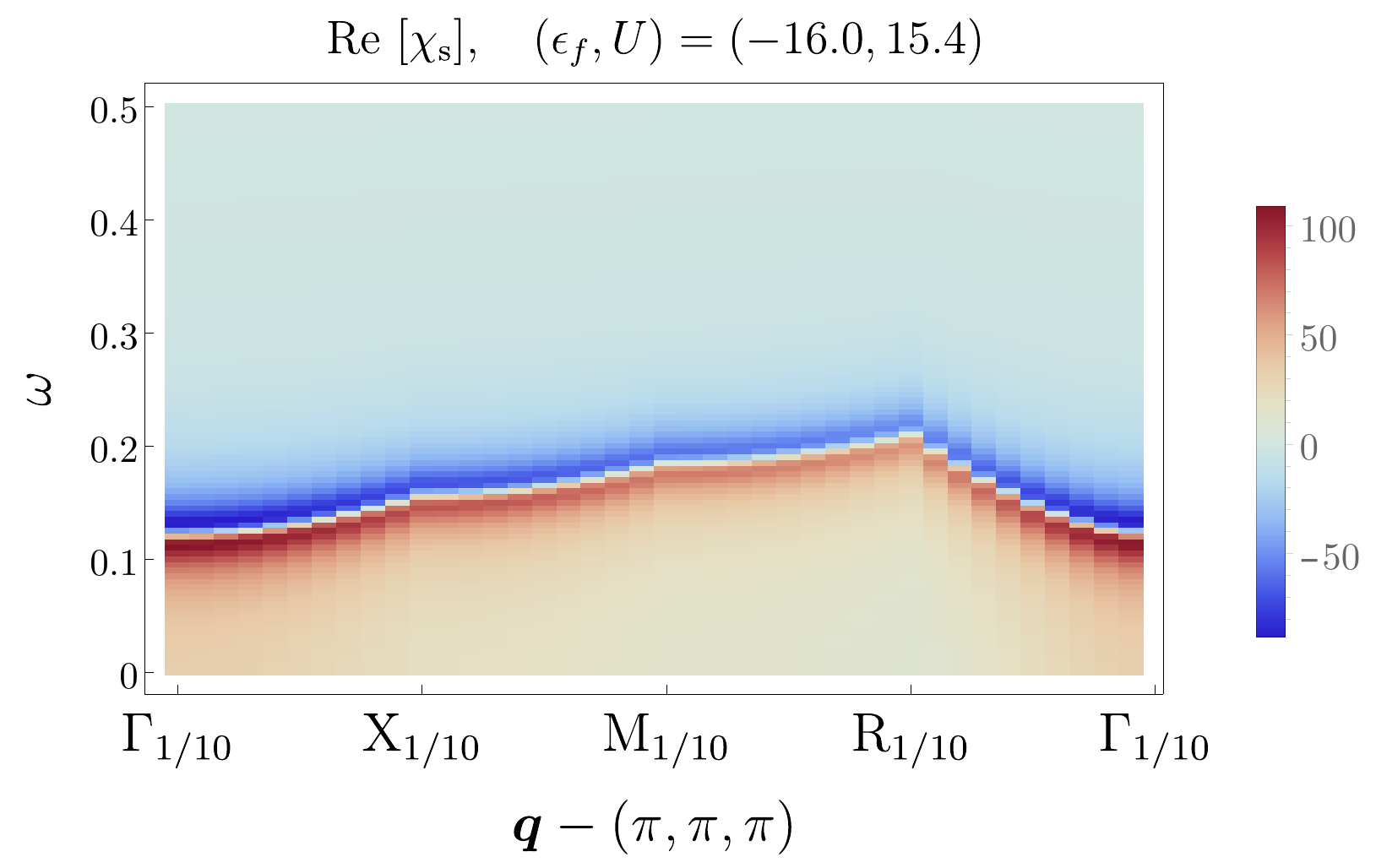}
\includegraphics[width=0.24\textwidth]{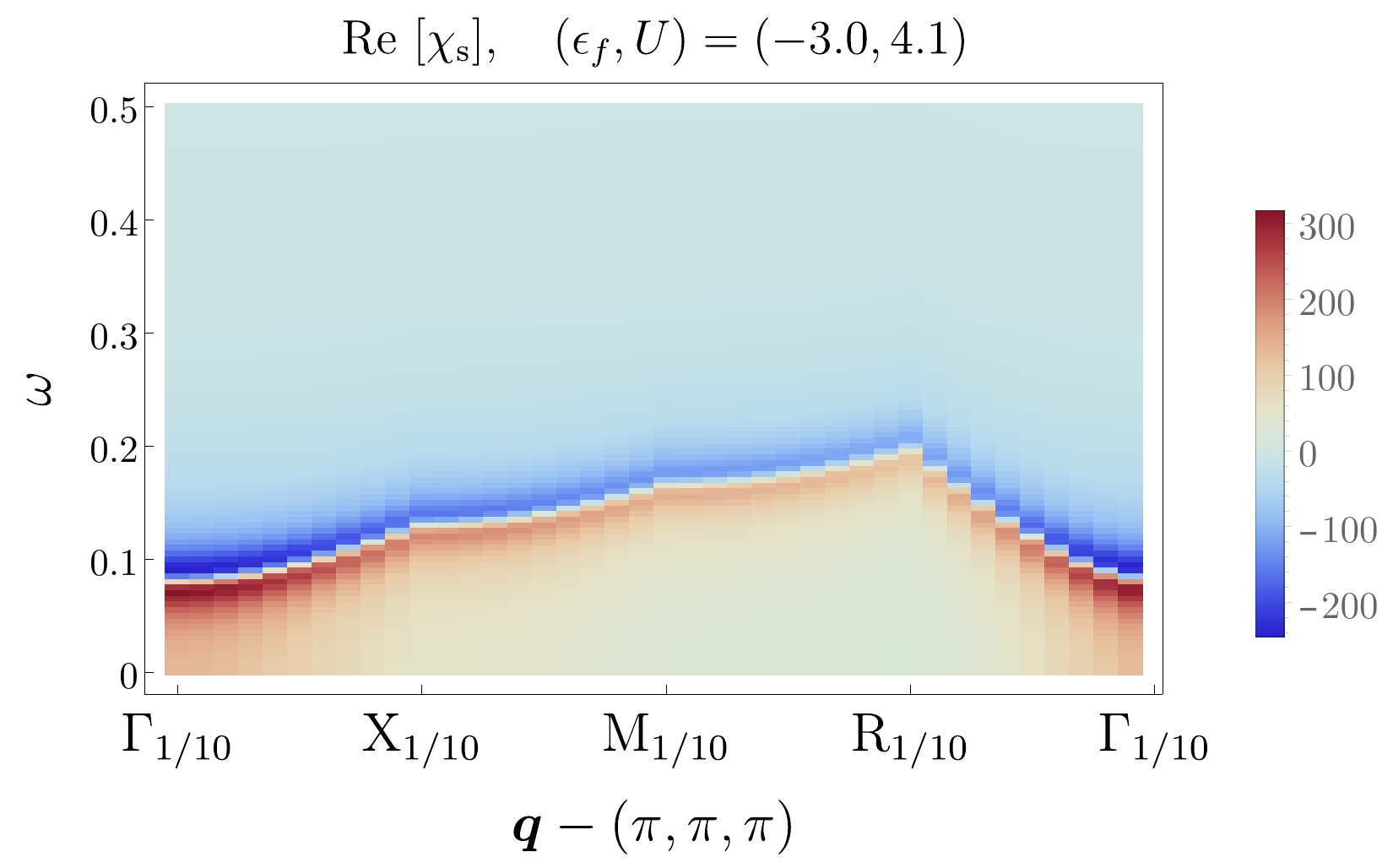}
\includegraphics[width=0.24\textwidth]{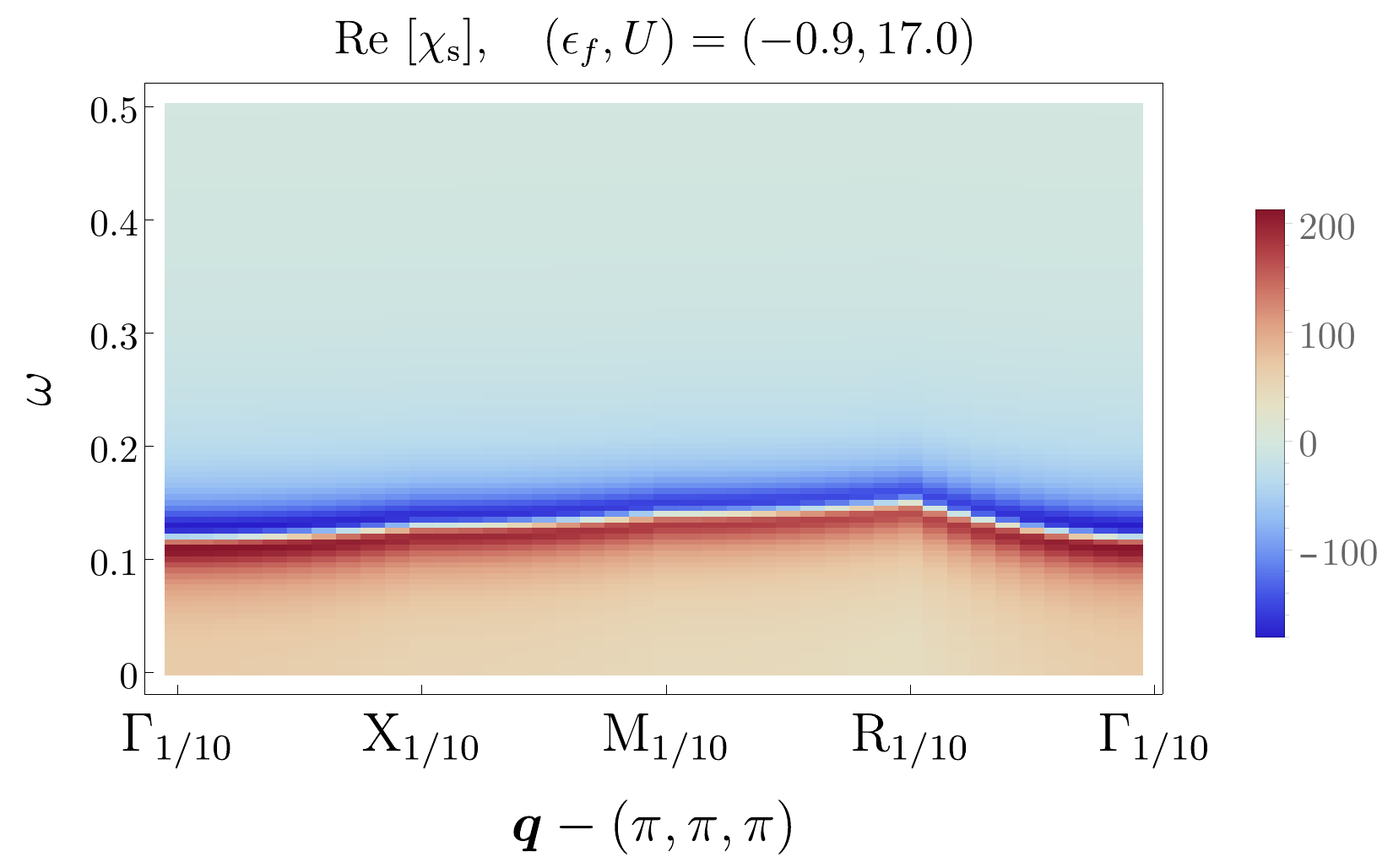}
\caption{
Susceptibility scan nearby $\bs Q=(\pi,\pi,\pi)$ at points close to the instability line in the paramagnetic phase.
The labels for reciprocal points are defined as $\Gamma_{1/10}=(0,0,0)$, $\textrm{X}_{1/10}=(\pi/10,0,0)$, $\textrm{M}_{1/10}=(\pi/10,\pi/10,0)$, and $\textrm{R}_{1/10}=(\pi/10,\pi/10,\pi/10)$.
$\Gamma_{1/10}=(0,0,0)$ requires the least energy ($\omega$) for the transition, which is another indicator of spontaneous ordering with $\bs Q=(\pi,\pi,\pi)$
}
\label{fig:3ntvsrltytg}
\end{figure}

		\begin{figure}[h]
			\centering
			\includegraphics[width=0.56\textwidth]{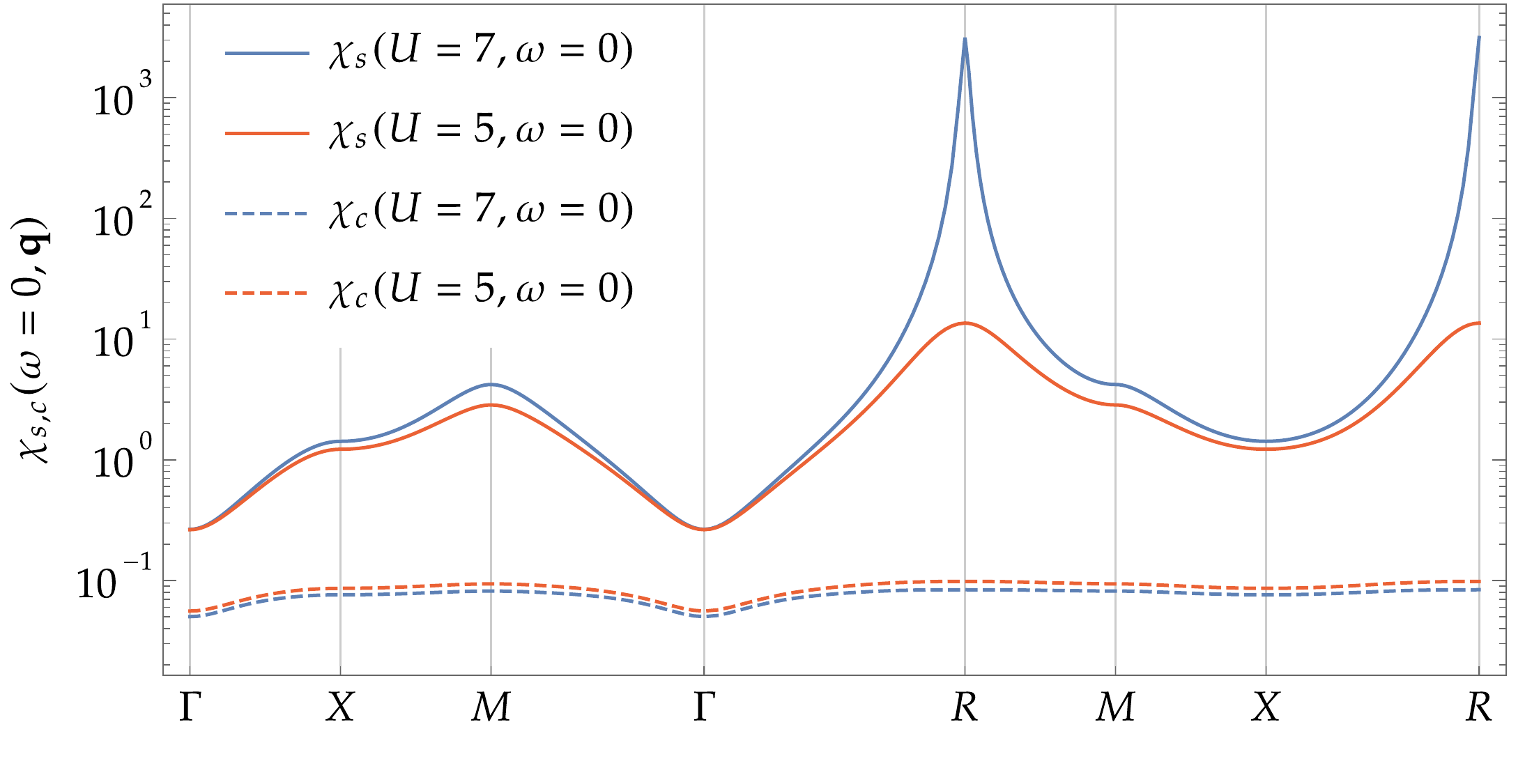}
			\caption{Spin and charge susceptibility at $\omega=0$ for $\epsilon_f=-1.25$ on the high symmetry path $\Gamma-\mbox{X}-\mbox{M}-\Gamma-\mbox{R}-\mbox{M}-\mbox{X}-\mbox{R}$,. The leading Fermi surface instability of the paramagnet is indeed magnetic with ordering vector $\bs Q=(\pi,\pi,\pi)$. There is no sign of incommensurate- or charge order.}
			\label{fig:5nwcaweh}
		\end{figure}

		\newpage
		\subsection{Detailed derivation of effective the mean field band structure}
		In this part, we derive the 8 by 8 matrix $\mathcal{H}_{\boldsymbol{k}, \bs Q}^{8 \times 8}$ for the magnetic ordering
		vector $\boldsymbol{Q}$ such that $2 \boldsymbol{Q} \equiv  \boldsymbol{G}$ (i.e. $\bs Q \in {(\pi,\pi,\pi),(\pi,\pi,0),(\pi,0,0)}$), where $(G_{x}, G_{y}, G_{y})=\boldsymbol{G}$ is a reciprocal lattice vector of the system.
		The original fermionic operators $\widetilde{\bs{f}}_i $ in slave boson formalism are represented by
		
		\begin{align}
		\begin{pmatrix}
		\tilde{f}_{i,\uparrow}\\ \tilde{f}_{i,\downarrow} 
		\end{pmatrix} 
		=
		z_{i}^T 
		\begin{pmatrix}
		f_{i,\uparrow}\\ f_{i, \downarrow}
		\end{pmatrix}, \quad
		z_{i} =e_i^\dagger ~L^\nodag_iM^\nodag_iR^\nodag_i\frac{1}{2} \left(p^\nodag_{i,0}\tau^0+\sum_{\alpha=1}^3 p^\nodag_{i,\alpha} \tau^\alpha\right) +\frac{1}{2} \left(p^\dagger_{i,0}\tau^0-\sum_{\alpha=1}^3 p^\dagger_{i,\alpha} \tau^\alpha\right) L^\nodag_iM^\nodag_iR^\nodag_i ~d_i, \label{eq:z_0}
		\end{align}
		with
		\begin{align}
		L_i^\nodag &= \left((1-d_i^\dagger d^\nodag_i)\tau^0-\frac{1}{2}\left(p^\dagger_{i,0}\tau^0+\sum_{\alpha=1}^3 p^\dagger_{i,\alpha} \tau^\alpha\right)  \left(p^\nodag_{i,0}\tau^0+\sum_{\alpha'=1}^3 p^\nodag_{i,\alpha'} \tau^{\alpha'}\right)\right)^{-1/2}, \\
		M_i^\nodag &= \left(1 + d_i^\dagger d_i^\nodag +e_i^\dagger e_i^\nodag +\sum\limits_{\alpha} p^\dagger_{i,\alpha} p_{i,\alpha}^\nodag \right)^{1/2}, \\
		R_i^\nodag &= \left((1-e_i^\dagger e_i^\nodag)\tau^0-\frac{1}{2}\left(p^\dagger_{i,0}\tau^0-\sum_{\alpha=1}^3 p^\dagger_{i,\alpha} \tau^\alpha\right)  \left(p^\nodag_{i,0}\tau^0-\sum_{\alpha'=1}^3 p^\nodag_{i,\alpha'} \tau^{\alpha'}\right)\right)^{-1/2}.
		\end{align}
		The Hilbert space defined by the slave particle operators has to be projected on to the physical Hilbert space by the application of constraints	
		\begin{equation}
		\begin{split}
		e^{\dagger}_{i} e_{i} +d^{\dagger}_{i} d_{i} + \sum_{\tilde{\alpha}=0}^3 p_{i,\tilde{\alpha}}^{\dagger} p_{i,\tilde{\alpha}}^\nodag  
		&= 1 \ ,
		\\
		\sum_{\tilde{\alpha}=0}^3 p_{i,\tilde{\alpha}}^{\dagger} p_{i,\tilde{\alpha}}^\nodag + 2d^{\dagger}_{i} d_{i}^\nodag  
		&= \sum_{\sigma} f_{i,\sigma}^{\dagger} f_{i,\sigma}^\nodag  \ ,
		\\
		p_{i,0}^{\dagger} \boldsymbol{p}_{i} ^\nodag + \boldsymbol{p}^{\dagger}_{i} p_{i,0}^\nodag  - \textrm{i} \boldsymbol{p}^{\dagger}_{i} \times \boldsymbol{p}_{i}^\nodag  
		&= \sum_{\sigma,\sigma'} \boldsymbol{\tau}_{\sigma \sigma'}^\nodag  f_{i,\sigma'}^{\dagger} f_{i,\sigma}^\nodag  \ ,
		\label{eq:3htrbwrb}
		\end{split}
		\end{equation}
		implying, respectively, that (i) the $f$ orbital at each site is empty, singly, or doubly occupied, (ii) the particle number as well as (iii) the spin operator of the $f$ electrons matches the one prescribed by the bosonic operators.
		These physical constraints are inserted in the Lagrangian of the system via five Lagrange multipliers $i\alpha_{i}$, $i\beta_{i,0}$, and $i\beta_{i,\alpha}$~\cite{PAM_Wuerzburg}.
	
		Using the mean field ansatz for the bosonic fields defined in Eq.~\eqref{mfansatz}, we find
		\begin{equation} 
		z_i\longrightarrow
		\begin{pmatrix}
		\mathcal{Z}_+ & \mathcal{Z}_- e^{-i\phi_i}\\
		\mathcal{Z}_- e^{i\phi_i} & \mathcal{Z}_+
		\end{pmatrix}
		\quad \text{with} \quad \mathcal{Z}_\pm= \frac{z_+ \pm z_-}{2}
		\quad \text{and} \quad  \phi_i=\boldsymbol{Q} \cdot \boldsymbol r_i .
		\label{eq:3pasdcb}
		\end{equation}

		We impose the constraint $1=e^2+d^2+p_0^2+p^2$, by directly substituting $e$, which is formally equivalent to integrating out the corresponding projector, yielding
		\begin{equation} 
		z_{\pm}
		=\frac{p_0(e+d)\pm p(e-d)}{\sqrt{2\left[1-d^2-(p_0\pm p)^2/2\right] \left[1-e^2-(p_0\mp p)^2/2\right]}}
		=\frac{p_0\left(\sqrt{1-p_0^2- p^2-d^2}+d \right)\pm  p \left(\sqrt{1-p_0^2- p^2-d^2}-d \right)}{\sqrt{2\left(1-d^2-(p_0\pm| p|)^2/2\right)\left(p_0^2+ p^2+d^2-(p_0\mp  p)^2/2\right)}}.
		\end{equation}
		Note that we could in principle have substituted any other bosonic field. 
		We apply the slave boson mean field ansatz to the Hamiltonian given in Eq.~\eqref{eq:1}
		\begin{equation}
		\label{app:Hamiltonian}
		\begin{split}
		H 
		=&
		H_0 +
		H_{\text{hyb}} +
		H_{\text{int}} := H^{ff}_0 + H^{cc}_0+ H_\mu + H_{\text{hyb}} + H_{\text{int}}
		\end{split}
		\end{equation}
		with
		\begin{align}
		\label{app:Hamiltonian_details}
		\begin{split}
		H^{ff}_0 
		=&
		-\sum_{i,j}\left( \widetilde{\bs{f}}_i^{~\dagger} t_{ij}^f\widetilde{\bs{f}}_j^\nodag + \text{H.c.} \right) \longrightarrow
		-\sum_{i,j} \left( \bs{f}_i^{~\dagger} z_i^* t_{ij}^f z_j^*\bs{f}_j^\nodag + \text{H.c.} \right)
		\\
		H^{cc}_0 
		=&
		-\sum_{i,j}\left( \bs c_i^{~\dagger} t_{ij}^d\bs c_j^\nodag + \text{H.c.} \right)
		\longrightarrow
		-\sum_{i,j}\left( \bs c_i^{~\dagger} t_{ij}^d\bs c_j^\nodag + \text{H.c.} \right)
		\\
		H_\mu=& \sum_{i} \widetilde{\bs{f}}_i^{~\dagger} \epsilon_{f} \widetilde{\bs{f}}_i^\nodag 
		-\sum_{i}  \mu_{0}  \left(\widetilde{\bs{f}}_i^{~\dagger}\widetilde{\bs{f}}_i^\nodag +\bs c_i^{~\dagger} \bs c_i^\nodag\right)
		\longrightarrow
		\sum_{i} \bs{f}_i^{~\dagger} \epsilon_{f} \bs{f}_i^\nodag 
		-\sum_{i}  \mu_{0}  \left(\bs f_i^{~\dagger}\bs f_i^\nodag +\bs c_i^{~\dagger} \bs c_i^\nodag\right)
		\\
		H_{\text{hyb}}
		=&
		\sum_\alpha \sum_{\langle i,j \rangle_\alpha }\text{i}V\left( \widetilde{\bs{f}}_i^{~\dagger} \tau^\alpha \bs c^\nodag_j + \bs c^\dagger_i \tau^\alpha \widetilde{\bs{f}}_j^\nodag + \text{H.c.} \right)
		\longrightarrow 
		\sum_\alpha \sum_{\langle i,j \rangle_\alpha }\text{i}V\left( \bs{f}_i^{~\dagger} z_i^* \tau^\alpha \bs c^\nodag_j + \bs c^\dagger_i \tau^\alpha z_j^* \bs{f}_j^\nodag + \text{H.c.} \right)
		\\
		H_{\text{int}} =& ~U \sum_{i} \tilde{f}_{i,\uparrow}^{\dagger} \tilde{f}_{i,\uparrow} 
		\tilde{f}_{i,\downarrow}^{\dagger} \tilde{f}_{i,\downarrow}
		\longrightarrow
		U \sum_{i} d^2.
		\end{split} 
		\end{align}
		The mean field Lagrangian $L$ for Eq.~\eqref{app:Hamiltonian} including the constraints reads
		\begin{equation}
		\begin{split}
		L=
		-\text{i} \sum_i \left(\bs f^\dagger_i \partial_t  \bs f^\nodag_i + \bs c^\dagger_i \partial_t \bs c^\nodag_i\right) + \tilde{H} 
		-\beta_{0} \sum_{i} \left(p_{0}^2+p^2+2d^2\right)
		-2\beta  \sum_{i} p_{0}^\nodag p^\nodag +\sum_i Ud^2
		\end{split}
		\end{equation}
		with
		\begin{equation}
		\tilde{H}:=H^{ff}_0+H^{cc}_0+H_\text{hyb}+H_\mu + \sum_{i} 
		\bs f_i^\dagger
		\begin{pmatrix}
		\beta_0 & \beta e^{\textrm{i} \phi_{i}} \\
		\beta e^{-\textrm{i} \phi_{i}} & \beta_0
		\end{pmatrix}
		\bs f_i^\nodag \ .
		\end{equation}
		To integrate out the fermionic degrees of freedom via the path integral formalism \cite{PAM_Wuerzburg}, we need to transform $\tilde{H}$ into momentum space. To do so, we choose the basis
		\begin{equation}
		\bs \Phi_{\bs k}^\dagger =
		\begin{pmatrix}
		c_{\boldsymbol{k},\uparrow}^{\dagger} & 
		c_{\boldsymbol{k},\downarrow}^{\dagger} &
		c_{\boldsymbol{k}+\bs Q,\uparrow}^{\dagger} & 
		c_{\boldsymbol{k}+\bs Q,\downarrow}^{\dagger} &
		f_{\boldsymbol{k},\uparrow}^{\dagger} &
		f_{\boldsymbol{k},\downarrow}^{\dagger} &
		f_{\boldsymbol{k}+\bs Q,\uparrow}^{\dagger} &
		f_{\boldsymbol{k}+\bs Q,\downarrow}^{\dagger}
		\end{pmatrix}
		\quad \text{with} \quad \tilde{H}= \sum_{\bs k} \bs \Phi_{\bs k}^\dagger \mathcal{H}_{\bs k, \bs Q}^{8 \times 8} \bs \Phi_{\bs k}^\nodag
		\ .
		\label{eq:5uktvd}
		\end{equation}
		The free energy is found to be
		\begin{equation}\label{app:freeenergy}
		\begin{gathered}
		\frac{F}{N}=-\frac TN\sum_{\nu = 1}^8\sum_{\boldsymbol k \in \text{BZ'}} \ln\left[1+e^{-\epsilon_{\boldsymbol k,\boldsymbol Q}^\nu/T}\right]
		+U d^2+n\mu_0-\beta_0(2d^2+ p^2 + p_0^2)-2\beta p_0 p,
		\end{gathered}
		\end{equation}
		where  $\epsilon^\nu_{\boldsymbol k,\boldsymbol Q}$ are the eigenvalues of $\mathcal{H}_{\bs k, \bs Q}^{8 \times 8}$, $N$ is the number of lattice sites and BZ' is the Brillouin zone of the corresponding magnetic unit cell. While the paramagnetic BZ contains $N$ discretized momenta, there are only $N/2$ in BZ', because the magnetic unit cell is doubled in size. The sum over $\bs k$ is always to be taken in BZ' in this chapter, which we will drop in the notation for readability.
		The values for the bosonic fields are given by the solving the saddle point equations for the free energy
		\begin{equation}
		\frac{\partial F}{\partial d}=\frac{\partial F}{\partial p_0}=\frac{\partial F}{\partial p}=\frac{\partial F}{\partial \beta_0}=\frac{\partial F}{\partial \beta}=\frac{\partial F}{\partial \mu_0}=0.
		\end{equation}
		
		In the following, we provide a detailed derivation of $\mathcal{H}_{\bs k, \bs Q}^{8 \times 8}$.
		Since $H_0^{cc}$ does not change under the slave-bosonic renormalization, one simply gets
		\begin{equation}
		\begin{split}
		H_0^{cc} = 
		\sum_{\boldsymbol{k}}
		\begin{pmatrix}
		c_{\boldsymbol{k},\uparrow}^{\dagger} &
		c_{\boldsymbol{k},\downarrow}^{\dagger} &
		c_{\boldsymbol{k}+\bs Q,\uparrow}^{\dagger} &
		c_{\boldsymbol{k}+\bs Q,\downarrow}^{\dagger}
		\end{pmatrix}
		\mathcal{H}^{cc}_{t} 
		\begin{pmatrix}
		c_{\boldsymbol{k},\uparrow} \\
		c_{\boldsymbol{k},\downarrow} \\
		c_{\boldsymbol{k}+\bs Q,\uparrow} \\
		c_{\boldsymbol{k}+\bs Q,\downarrow} 
		\end{pmatrix}, \quad 
		\\
		\mathcal{H}^{cc}_{t}
		=
		\begin{pmatrix}
		\xi^{c}_{\boldsymbol{k}} & 0 & 0 & 0\\
		0 & \xi^{c}_{\boldsymbol{k}} & 0 & 0\\
		0 & 0 & \xi^{c}_{\boldsymbol{k}+\bs Q} & 0\\
		0 & 0 & 0 & \xi^{c}_{\boldsymbol{k}+\bs Q}\\
		\end{pmatrix}
		\ , \quad
		\xi^{c}_{\boldsymbol{k}} := \frac{1}{N} \sum_{\bs r_{ij}} t^{c}_{ij} e^{-\textrm{i} \boldsymbol{k} \boldsymbol{r}_{ij}} +\text{H.c.}\ .
		\end{split}
		\label{eq:6yudscsb}
		\end{equation}
		where $\bs r_{ij}:= \bs r_i-\bs r_j$ connects the sites $i$ and $j$. Note, that the matrix $\mathcal{H}^{cc}$ as been symmetrized to the extended basis, i.e. there appear terms like $\xi^{c}_{\boldsymbol{k}+\bs Q}$ additionally to the expected terms
		$\xi^{c}_{\boldsymbol{k}}$. This is necessary to obtain a Bloch form for the Hamilationan matrix $\mathcal{H}$ and find the correct results 
		via the $  \boldsymbol{k}$-integral in the smaller magnetic BZ'. For the same reason, the following terms will also be symmetrized in the 
		same fashion.
		
		\textit{Hybridization term} ---
		The hybridization term can be written as
		\begin{align}
		H^{\textrm{hyb}}&=\textrm{i}V \sum_{\alpha} \sum_{\langle i,j \rangle_{\alpha}}
		\left(
		\boldsymbol{f}^{\dagger}_i z^\nodag_i \tau^{\alpha} \bs c_j^\nodag
		-\boldsymbol{f}^{\dagger}_j z^\nodag_j \tau^\alpha \bs c_i^\nodag +\text{H.c.} 
		\right)
		\ .
		\end{align}
		we find
		\begin{align}
		&\phantom{=,} \textrm{i}V \sum_{\alpha} \sum_{\langle i,j \rangle_{\alpha}}
		\left(
		\boldsymbol{f}^{\dagger}_i z^{\nodag}_i \tau^{\alpha} \bs c_j^\nodag 
		-\boldsymbol{f}^{\dagger}_j z^{\nodag}_j \tau^\alpha \bs c_i^\nodag
		\right) \nonumber \\
		&= \textrm{i} \frac{V}{N/2} \sum_{\boldsymbol{k},\boldsymbol{k}'} \sum_{i,\alpha}
		e^{-\textrm{i} \boldsymbol{k} \boldsymbol{r}_{i}}
		e^{\textrm{i} \boldsymbol{k}' (\boldsymbol{r}_{i}+\hat{\boldsymbol{r}}_{\alpha})} 
		\bs f^\dagger_{\bs k}
		\begin{pmatrix}
		\mathcal{Z}_+ & \mathcal{Z}_- e^{\textrm{i} \phi_i} \\
		\mathcal{Z}_- e^{-\textrm{i} \phi_i} & \mathcal{Z}_+
		\end{pmatrix}
		\tau^\alpha
		\bs c^\nodag_{\bs k'} \nonumber \\
		&-\textrm{i} \frac{V}{N/2} \sum_{\boldsymbol{k},\boldsymbol{k}'} \sum_{j,\alpha}
		e^{-\textrm{i} \boldsymbol{k} \boldsymbol{r}_{j}}
		e^{\textrm{i} \boldsymbol{k}' (\boldsymbol{r}_{j}-\hat{\boldsymbol{r}}_{\alpha})}
		\bs f^\dagger_{\bs k}
		\begin{pmatrix}
		\mathcal{Z}_+ & \mathcal{Z}_- e^{\textrm{i} \phi_j} \\
		\mathcal{Z}_- e^{-\textrm{i} \phi_j} & \mathcal{Z}_+
		\end{pmatrix}
		\tau^\alpha
		\bs c^\nodag_{\bs k'} \nonumber \\
		&=
		\textrm{i} \frac{V}{N/2} \sum_{\boldsymbol{k},\boldsymbol{k}'} \sum_{i,\alpha}
		\bs f^\dagger_{\bs k}
		\begin{pmatrix}
		\mathcal{Z}_+(
		e^{-\textrm{i} \boldsymbol{k} \boldsymbol{r}_{i}}
		e^{\textrm{i} \boldsymbol{k}' (\boldsymbol{r}_{i}+\hat{\boldsymbol{r}}_{\alpha})}
		-e^{-\textrm{i} \boldsymbol{k} (\boldsymbol{r}_{i}+\hat{\boldsymbol{r}}_{\alpha})}
		e^{\textrm{i} \boldsymbol{k}' \boldsymbol{r}_{i}}
		)  & (2,1)_{(\bs Q \rightarrow -\bs Q)} \\
		\mathcal{Z}_-(
		e^{-\textrm{i} \boldsymbol{k} \boldsymbol{r}_{i}}
		e^{\textrm{i} \boldsymbol{k}' (\boldsymbol{r}_{i}+\hat{\boldsymbol{r}}_{\alpha})}
		e^{-\textrm{i} \bs Q \boldsymbol{r}_i}
		-e^{-\textrm{i} \boldsymbol{k} (\boldsymbol{r}_{i}+\hat{\boldsymbol{r}}_{\alpha})}
		e^{\textrm{i} \boldsymbol{k}' \boldsymbol{r}_{i}}
		e^{-\textrm{i} \bs Q (\boldsymbol{r}_i+\hat{\boldsymbol{r}}_\alpha)}
		)
		& (1,1)
		\end{pmatrix}
		\tau^{\alpha}
		\bs c^\nodag_{\bs k'} \nonumber \\
		&=
		\textrm{i} \frac{V}{N/2} \sum_{\boldsymbol{k},\boldsymbol{k}'} \sum_{i,\alpha}
		\bs f^\dagger_{\bs k}
		\begin{pmatrix}
		\mathcal{Z}_+\,
		e^{\textrm{i} \boldsymbol{r}_i (-\boldsymbol{k}+\boldsymbol{k}')}
		(e^{\textrm{i} \boldsymbol{k}' \hat{\boldsymbol{r}}_\alpha} - e^{-\textrm{i} \boldsymbol{k} \hat{\boldsymbol{r}}_\alpha})
		& (2,1)_{(\bs Q \rightarrow -\bs Q)} \\
		\mathcal{Z}_-\,
		e^{\textrm{i} \boldsymbol{r}_i (-\boldsymbol{k}+\boldsymbol{k}'-\bs Q)}
		(e^{\textrm{i} \boldsymbol{k}' \hat{\boldsymbol{r}}_\alpha} - e^{-\textrm{i} \boldsymbol{k} \hat{\boldsymbol{r}}_\alpha} e^{-\textrm{i} \bs Q \hat{\boldsymbol{r}}_\alpha})
		& (1,1)
		\end{pmatrix}
		\tau^{\alpha}
		\bs c^\nodag_{\bs k'} \nonumber \\
		&=2
		\textrm{i} V \sum_{\boldsymbol{k},\boldsymbol{k}'} \sum_{\alpha}
		\bs f^\dagger_{\bs k}
		\begin{pmatrix}
		\mathcal{Z}_+\,
		\delta_{\boldsymbol{k}',\boldsymbol{k}}
		(e^{\textrm{i} \boldsymbol{k}' \hat{\boldsymbol{r}}_\alpha} - e^{-\textrm{i} \boldsymbol{k} \hat{\boldsymbol{r}}_\alpha})
		& (2,1)_{(\bs Q \rightarrow -\bs Q)} \\
		\mathcal{Z}_-\,
		\delta_{\boldsymbol{k}',\boldsymbol{k}+\bs Q}
		(e^{\textrm{i} \boldsymbol{k}' \hat{\boldsymbol{r}}_\alpha} - e^{-\textrm{i} \boldsymbol{k} \hat{\boldsymbol{r}}_\alpha} e^{\textrm{i} \bs Q \hat{\boldsymbol{r}}_\alpha})
		& (1,1)
		\end{pmatrix}
		\tau^{\alpha}
		\bs c^\nodag_{\bs k'} \nonumber \\
		&=2
		\textrm{i} V \sum_{\boldsymbol{k},\boldsymbol{k}'} \sum_{\alpha}
		\bs f^\dagger_{\bs k}
		\begin{pmatrix}
		\mathcal{Z}_+\,
		\delta_{\boldsymbol{k}',\boldsymbol{k}}\,
		2 \textrm{i} \sin (\boldsymbol{k} \hat{\boldsymbol{r}}_\alpha)
		& 
		\mathcal{Z}_-\,
		\delta_{\boldsymbol{k}',\boldsymbol{k}-\bs Q}\,
		2 \textrm{i} \sin ((\boldsymbol{k}-\bs Q)\hat{\boldsymbol{r}}_{\alpha}) \\
		\mathcal{Z}_-\,
		\delta_{\boldsymbol{k}',\boldsymbol{k}+\bs Q}\,
		2 \textrm{i} \sin ((\boldsymbol{k}+\bs Q)\hat{\boldsymbol{r}}_{\alpha})
		& 
		\mathcal{Z}_+\,
		\delta_{\boldsymbol{k}',\boldsymbol{k}}\,
		2 \textrm{i} \sin (\boldsymbol{k} \hat{\boldsymbol{r}}_\alpha)
		\end{pmatrix}
		\tau^{\alpha}
		\bs c^\nodag_{\bs k'} \nonumber \\
		&=
		- V \sum_{\boldsymbol{k}} \sum_{\alpha}
		\Bigg[
		\bs f^\dagger_{\bs k}
		\mathcal{Z}_+ 2\sin (\boldsymbol{k} \hat{\boldsymbol{r}}_\alpha)\, \tau^\alpha 
		\bs c^\nodag_{\bs k}
		+
		\bs f^\dagger_{\bs k+\bs Q}
		\mathcal{Z}_+ 2\sin ((\boldsymbol{k}+\bs Q) \hat{\boldsymbol{r}}_\alpha)\, \tau^\alpha 
		\bs c^\nodag_{\bs k+\bs Q}
		\nonumber 
		\\&\phantom{\frac{2V}{2} \sum_{\boldsymbol{k},\boldsymbol{k}'} \sum_{\alpha}+}+
		\bs f^\dagger_{\bs k+\bs Q}
		\mathcal{Z}_- 2\sin (\boldsymbol{k}\hat{\boldsymbol{r}}_{\alpha})\, \tau^{+}\, \tau^\alpha
		\bs c^\nodag_{\bs k}
		+
		\bs f^\dagger_{\bs k}
		\mathcal{Z}_- 2\sin ((\boldsymbol{k}-\bs Q)\hat{\boldsymbol{r}}_{\alpha})\, \tau^{+}\, \tau^\alpha
		\bs c^\nodag_{\bs k-\bs Q}
		\nonumber
		\\&\phantom{\frac{2V}{2} \sum_{\boldsymbol{k},\boldsymbol{k}'} \sum_{\alpha}+}+
		\bs f^\dagger_{\bs k-\bs Q}
		\mathcal{Z}_- 2\sin (\boldsymbol{k}\hat{\boldsymbol{r}}_{\alpha})\, \tau^{-}\, \tau^\alpha
		\bs c^\nodag_{\bs k}
		+
		\bs f^\dagger_{\bs k}
		\mathcal{Z}_- 2\sin ((\boldsymbol{k}+\bs Q)\hat{\boldsymbol{r}}_{\alpha})\, \tau^{-}\, \tau^\alpha
		\bs c^\nodag_{\bs k+\bs Q}
		\Bigg]
		\ . \nonumber
		\label{eq:14dcae}
		\end{align} 
		Here, $(2,1)_{(\bs Q \rightarrow -\bs Q)}$ indicates the same object as the $(2,1)$-component in the same matrix with $\bs Q$ changed into $-\bs Q$.
		Similarly, $(1,1)$ is the same one as the $(1,1)$-component.
		Further we define $\tau^{\pm} := (\tau^{1} \pm \textrm{i} \tau^{2})/2$ is introduced for convenience of notations.
		Since $\bs Q= 2\bs G$, $\boldsymbol{k}-\bs Q$ is equivalent to $\boldsymbol{k}+\bs Q$ up to reciprocal space symmetry. 
		Therefore, we find
		\begin{equation}
		\begin{split}
		&\textrm{i}V \sum_{\alpha} \sum_{\langle i,j \rangle_{\alpha}}
		\left(
		\boldsymbol{f}^{\dagger}_i \underline{z}^{\dagger}_i \tau^{\alpha} \boldsymbol{c}_j 
		-\boldsymbol{f}^{\dagger}_j \underline{z}^{\dagger}_j \tau^\alpha \boldsymbol{c}_i
		\right) \\
		&
		=- V \sum_{\boldsymbol{k}} \sum_{\alpha}
		\Bigg[
		\mathcal{Z}_+ 2\sin (\boldsymbol{k} \hat{\boldsymbol{r}}_\alpha)
		\bs f^\dagger_{\bs k}
		\tau^\alpha 
		\bs c^\nodag_{\bs k}
		+
		\mathcal{Z}_+ 2\sin ((\boldsymbol{k}+\bs Q) \hat{\boldsymbol{r}}_\alpha)
		\bs f^\dagger_{\bs k+\bs Q}
		\tau^\alpha 
		\bs c^\nodag_{\bs k+\bs Q}
		\\&\phantom{\frac{2V}{2} \sum_{\boldsymbol{k},\boldsymbol{k}'} \sum_{\alpha}}+
		\mathcal{Z}_- 2\sin (\boldsymbol{k}\hat{\boldsymbol{r}}_{\alpha})
		\bs f^\dagger_{\bs k+\bs Q}
		\tau^{1} \tau^\alpha
		\bs c^\nodag_{\bs k}
		+
		\mathcal{Z}_- 2\sin ((\boldsymbol{k}+\bs Q)\hat{\boldsymbol{r}}_{\alpha})
		\bs f^\dagger_{\bs k}
		\tau^{1} \tau^\alpha
		\bs c^\nodag_{\bs k+\bs Q}
		\Bigg] \ .
		\label{eq:14dcae}
		\end{split}
		\end{equation} 
		Eq.~\eqref{eq:14dcae} can be represented with $4 \times 4$ matrix $\mathcal{H}^{fc}$ in the chosen basis as
		\begin{equation}
		\sum_{\alpha} \sum_{\langle i,j \rangle _{\alpha}}
		\left[
		\tilde{\boldsymbol{f}}^{\dagger}_{i} (\textrm{i}V \tau^{\alpha}) \boldsymbol{c}_{j} 
		-\tilde{\boldsymbol{f}}^{\dagger}_{j} (\textrm{i}V \tau^{\alpha}) \boldsymbol{c}_{i}
		\right] 
		=
		\sum_{\boldsymbol{k}}
		\begin{pmatrix}
		f_{\boldsymbol{k},\uparrow}^{\dagger} &
		f_{\boldsymbol{k},\downarrow}^{\dagger} &
		f_{\boldsymbol{k}+\bs Q,\uparrow}^{\dagger} &
		f_{\boldsymbol{k}+\bs Q,\downarrow}^{\dagger}
		\end{pmatrix}
		\mathcal{H}^{fc} 
		\begin{pmatrix}
		c_{\boldsymbol{k},\uparrow} \\
		c_{\boldsymbol{k},\downarrow} \\
		c_{\boldsymbol{k}+\bs Q,\uparrow} \\
		c_{\boldsymbol{k}+\bs Q,\downarrow} 
		\end{pmatrix}
		\end{equation}
		with
		\begin{equation}
		\mathcal{H}^{fc} := 
		-V 
		\begin{pmatrix}
		\mathcal{Z}_+ s_{z,\bs k} & \mathcal{Z}_+ (s_{x,\bs k} -  \textrm{i}s_{y,\bs k}) & \mathcal{Z}_-(s_{x,\bs k+\bs Q} +  \textrm{i}s_{y,\bs k+ \bs Q})& -\mathcal{Z}_- s_{z,\bs k+\bs Q} \\
		\mathcal{Z}_+ (s_{x,\bs k}+  \textrm{i}s_{y,\bs k}) & -\mathcal{Z}_+ s_{z,\bs k} & \mathcal{Z}_- s_{z,\bs k+\bs Q} & \mathcal{Z}_-(s_{x,\bs k+\bs Q} -  \textrm{i} s_{y,\bs k+\bs Q})\\
		\mathcal{Z}_-(s_{x,\bs k} + \textrm{i} s_{y,\bs k})& -\mathcal{Z}_- s_{z,\bs k} & \mathcal{Z}_+ s_{z,\bs k+\bs Q} & \mathcal{Z}_+ (s_{x,\bs k+\bs Q} -  \textrm{i} s_{y,\bs k+\bs Q}) \\
		\mathcal{Z}_- s_{z,\bs k} & \mathcal{Z}_-(s_{x,\bs k} -  \textrm{i}s_{y,\bs k}) & \mathcal{Z}_+ (s_{x,\bs k+\bs Q} +  \textrm{i}s_{y,\bs k+\bs Q}) & -\mathcal{Z}_+ s_{z,\bs k+\bs Q}
		\end{pmatrix} \ ,
		\label{eq:14kjrtew}
		\end{equation}
		where we have defined 
		\begin{equation}
		s_{\alpha,\bs k} := 2\sin(\hat{\bs r}_\alpha \bs k).
		\end{equation}
		To account the hermitian conjugate part of Eq.~\eqref{eq:14dcae} in $\mathcal{H}^{\textrm{hyb}}$, we define $\mathcal{H}^{cf} := \big( \mathcal{H}^{fc} \big)^{\dagger}$. 
		
		\textit{Hopping terms within $f$-band} ---
		Below, $\sum_{i,j}$ indicates the sum over pairs of lattice sites of $i$ and $j$ with hopping amplitude $t^{f}_{\boldsymbol_{ij}}$ and $\boldsymbol{r}_{i}-\boldsymbol{r}_{j}:=\boldsymbol{r}_{ij}$
		\begin{align}
		\begin{split}
		\sum_{i,j}
		\tilde{\boldsymbol{f}}^{\dagger}_{i} t^{f}_{\boldsymbol_{ij}} \tilde{\boldsymbol{f}}_{j}
		&=
		\sum_{i,j} t^{f}_{\boldsymbol_{ij}}
		\boldsymbol{f}^{\dagger}_i
		\begin{pmatrix}
		\mathcal{Z}_+ & \mathcal{Z}_-\, e^{\textrm{i} \phi_i} \\
		\mathcal{Z}_-\, e^{-\textrm{i} \phi_i} & \mathcal{Z}_+
		\end{pmatrix}
		\begin{pmatrix}
		\mathcal{Z}_+ & \mathcal{Z}_-\, e^{\textrm{i} \phi_j} \\
		\mathcal{Z}_-\, e^{-\textrm{i} \phi_j} & \mathcal{Z}_+
		\end{pmatrix}
		\boldsymbol{f}_{j} \\
		&=
		\sum_{i,j} t^{f}_{\boldsymbol_{ij}}
		\boldsymbol{f}^{\dagger}_i
		\begin{pmatrix}
		\mathcal{Z}_+^2 + \mathcal{Z}_-^2\, e^{\textrm{i} \phi_i - \textrm{i} \phi_j} &
		\mathcal{Z}_+ \mathcal{Z}_-(e^{\textrm{i}\phi_i} + e^{\textrm{i} \phi_j}) \\
		\mathcal{Z}_+ \mathcal{Z}_- (e^{-\textrm{i}\phi_i} + e^{-\textrm{i} \phi_j}) &
		\mathcal{Z}_+^2 + \mathcal{Z}_-^2\, e^{-\textrm{i} \phi_i + \textrm{i} \phi_j}
		\end{pmatrix}
		\boldsymbol{f}_{j} \\
		&=
		\frac{1}{N/2} 
		\sum_{\boldsymbol{k},\boldsymbol{k}'} \sum_{i,j}  t^{f}_{\boldsymbol_{ij}}
		e^{-\textrm{i} \boldsymbol{k} \boldsymbol{r}_i} e^{\textrm{i} \boldsymbol{k}' (\bs r_i-\bs r_{ij})}
		\boldsymbol{f}^{\dagger}_{\boldsymbol{k}}
		\begin{pmatrix}
		\mathcal{Z}_+^2 + \mathcal{Z}_-^2\, e^{\textrm{i} \bs Q \boldsymbol{r}_{ij}} &
		\mathcal{Z}_+ \mathcal{Z}_-\, e^{\textrm{i} \bs Q \boldsymbol{r}_i} (1 + e^{-\textrm{i} \bs Q \boldsymbol{r}_{ij}}) \\
		\mathcal{Z}_+ \mathcal{Z}_-\, e^{-\textrm{i} \bs Q \boldsymbol{r}_i} (1 + e^{\textrm{i} \bs Q \boldsymbol{r}_{ij}}) &
		\mathcal{Z}_+^2 + \mathcal{Z}_-^2\, e^{-\textrm{i} \bs Q \boldsymbol{r}_{ij}}
		\end{pmatrix}
		\boldsymbol{f}_{\boldsymbol{k}'}\\
		&=
		2  \sum_{\boldsymbol{k},\boldsymbol{k}'} \sum_{\boldsymbol{r}_{ij}} t^{f}_{\boldsymbol_{ij}}
		e^{-\boldsymbol{i} \boldsymbol{k}' \boldsymbol{r}_{ij}}
		\boldsymbol{f}^{\dagger}_{\boldsymbol{k}}
		\begin{pmatrix}
		\delta_{\boldsymbol{k}',\boldsymbol{k}}\, (\mathcal{Z}_+^2 + \mathcal{Z}_-^2 e^{\textrm{i} \bs Q \boldsymbol{r}_{ij}}) &
		\delta_{\boldsymbol{k}',\boldsymbol{k}-\bs Q}\, \mathcal{Z}_+ \mathcal{Z}_- (1 + e^{-\textrm{i} \bs Q \boldsymbol{r}_{ij}}) \\
		\delta_{\boldsymbol{k},\boldsymbol{k}'-\bs Q}\, \mathcal{Z}_+ \mathcal{Z}_- 
		(1 + e^{\textrm{i} \bs Q \boldsymbol{r}_{ij}}) &
		\delta_{\boldsymbol{k},\boldsymbol{k}'}\, (\mathcal{Z}_+^2 + \mathcal{Z}_-^2 e^{-\textrm{i} \bs Q \boldsymbol{r}_{ij}})
		\end{pmatrix}
		\boldsymbol{f}_{\boldsymbol{k}'} \\
		&=2 \sum_{\boldsymbol{k}} \sum_{\boldsymbol{r}_{ij}} t^{f}_{\boldsymbol_{ij}}
		\Bigg	[
		e^{-\boldsymbol{i} \boldsymbol{k} \boldsymbol{r}_{ij}}
		\boldsymbol{f}^{\dagger}_{\boldsymbol{k}}
		\begin{pmatrix}
		\mathcal{Z}_+^2 + \mathcal{Z}_-^2 e^{\textrm{i} \bs Q \boldsymbol{r}_{ij}} & 0 \\
		0 & \mathcal{Z}_+^2 + \mathcal{Z}_-^2 e^{-\textrm{i} \bs Q \boldsymbol{r}_{ij}}
		\end{pmatrix}
		\boldsymbol{f}_{\boldsymbol{k}}
		\\
		&\phantom{2 \sum_{\boldsymbol{k}} \sum_{\boldsymbol{r}_{ij}} t^{f}_{\boldsymbol_{ij}} +}
		+
		\mathcal{Z}_+ \mathcal{Z}_-
		e^{\boldsymbol{i} (-\boldsymbol{k}+\bs Q) \boldsymbol{r}_{ij}}
		(1 + e^{-\textrm{i} \bs Q \boldsymbol{r}_{ij}})\,
		\boldsymbol{f}^{\dagger}_{\boldsymbol{k}}\,
		\tau^+
		\boldsymbol{f}_{\boldsymbol{k}-\boldsymbol Q} \\
		&\phantom{2 \sum_{\boldsymbol{k}} \sum_{\boldsymbol{r}_{ij}} t^{f}_{\boldsymbol_{ij}} +}
		+
		\mathcal{Z}_+ \mathcal{Z}_-
		e^{-\boldsymbol{i} (\boldsymbol{k}+\bs Q) \boldsymbol{r}_{ij}}
		(1 + e^{-\textrm{i} \bs Q \boldsymbol{r}_{ij}})\,
		\boldsymbol{f}^{\dagger}_{\boldsymbol{k}}\,
		\tau^-
		\boldsymbol{f}_{\boldsymbol{k}+\boldsymbol Q}
		\Bigg] \ 
		\\
		&=
		2\sum_{\boldsymbol{k}}
		\Bigg[
		\boldsymbol{f}^{\dagger}_{\boldsymbol{k}}
		\begin{pmatrix}
		\mathcal{Z}_+^2 \xi^f_{\boldsymbol{k}} + \mathcal{Z}_-^2 \xi^f_{\boldsymbol{k}-\bs Q} & 0 \\
		0 & \mathcal{Z}_+^2 \xi^f_{\boldsymbol{k}} + \mathcal{Z}_-^2 \xi^f_{\boldsymbol{k}+\bs Q}
		\end{pmatrix}
		\boldsymbol{f}_{\boldsymbol{k}}
		\\
		&\phantom{2\sum_{\bs k}+}
		+
		\mathcal{Z}_+ \mathcal{Z}_- (\xi^f_{\boldsymbol{k}-\bs Q}+\xi^f_{\boldsymbol{k}})\,
		\boldsymbol{f}^{\dagger}_{\boldsymbol{k}}\,
		\tau^+
		\boldsymbol{f}_{\boldsymbol{k}-\boldsymbol Q}
		+
		\mathcal{Z}_+ \mathcal{Z}_- 
		(\xi^f_{\boldsymbol{k}+\bs Q}+\xi^f_{\boldsymbol{k}})\,
		\boldsymbol{f}^{\dagger}_{\boldsymbol{k}}\,
		\tau^-
		\boldsymbol{f}_{\boldsymbol{k}+\boldsymbol Q} 
		\Bigg].\label{eq:fhopping}
		\end{split}
		\end{align}
		Above, reused the previous definition $\tau^\pm$ and we further define $\xi^{f}_{\boldsymbol{k}} := \sum_{\bs r_{ij}} t^{f}_{ij} e^{-\textrm{i} \boldsymbol{k} \bs r_{ij}}$. In order to obtain a Bloch form, we need to symmetrize the previous result to the basis defined in Eq.~\ref{eq:5uktvd} by replacing $(\ref{eq:fhopping})_{\bs k}\rightarrow \frac12\left((\ref{eq:fhopping})_{\bs k}+(\ref{eq:fhopping})_{\bs k + \bs Q} \right)$.
		Using once more the fact that $\boldsymbol{k}-\bs Q \equiv \boldsymbol{k}+\bs Q$, one finds
		\begin{align}
		\sum_{i,j}
		\tilde{\boldsymbol{f}}^{\dagger}_{i} t^{f}_{\boldsymbol_{ij}} \tilde{\boldsymbol{f}}_{j}
		=&
		\sum_{\boldsymbol{k}}
		\Bigg[
		\boldsymbol{f}^{\dagger}_{\boldsymbol{k}}
		\begin{pmatrix}
		\mathcal{Z}_+^2 \xi^f_{\boldsymbol{k}} + \mathcal{Z}_-^2 \xi^f_{\boldsymbol{k}+\bs Q} & 0 \\
		0 & \mathcal{Z}_+^2 \xi^f_{\boldsymbol{k}} + \mathcal{Z}_-^2 \xi^f_{\boldsymbol{k}-\bs Q}
		\end{pmatrix}
		\boldsymbol{f}_{\boldsymbol{k}}
		+
		\boldsymbol{f}^{\dagger}_{\boldsymbol{k}+\bs Q}
		\begin{pmatrix}
		\mathcal{Z}_+^2 \xi^f_{\boldsymbol{k}+\bs Q} + \mathcal{Z}_-^2 \xi^f_{\boldsymbol{k}} & 0 \\
		0 & \mathcal{Z}_+^2 \xi^f_{\boldsymbol{k}+\bs Q} + \mathcal{Z}_-^2 \xi^f_{\boldsymbol{k}}
		\end{pmatrix}
		\boldsymbol{f}_{\boldsymbol{k}+\bs Q}
		\nonumber \\
		&\phantom{\frac{1}{2}\sum_{\boldsymbol{k}}}
		+
		\mathcal{Z}_+ \mathcal{Z}_- (\xi^f_{\boldsymbol{k}+\bs Q}+\xi^f_{\boldsymbol{k}})
		(\boldsymbol{f}^{\dagger}_{\boldsymbol{k}}\, \tau^{1} \boldsymbol{f}_{\boldsymbol{k}+\bs Q}
		+
		\boldsymbol{f}^{\dagger}_{\boldsymbol{k}+\bs Q}\, \tau^{1} \boldsymbol{f}_{\boldsymbol{k}})
		\Bigg] \ .
		\label{eq:18jneewqbc}
		\end{align}
		Eq.~\ref{eq:18jneewqbc} can be represented with a $4\times 4$ matrix $\mathcal{H}^{ff}_{t}$ in the chosen basis as
		\begin{equation}
		\sum_{i,j}
		\tilde{\boldsymbol{f}}^{\dagger}_{i} t^{f}_{\boldsymbol_{ij}} \tilde{\boldsymbol{f}}_{j}
		=
		\sum_{\boldsymbol{k}}
		\begin{pmatrix}
		f^{\dagger}_{\boldsymbol{k},\uparrow} &
		f^{\dagger}_{\boldsymbol{k},\downarrow} &
		f^{\dagger}_{\boldsymbol{k}+\bs Q,\uparrow} &
		f^{\dagger}_{\boldsymbol{k}+\bs Q,\downarrow}
		\end{pmatrix}
		\mathcal{H}^{ff}
		\begin{pmatrix}
		f_{\boldsymbol{k},\uparrow} \\
		f_{\boldsymbol{k},\downarrow} \\
		f_{\boldsymbol{k}+\bs Q,\uparrow} \\
		f_{\boldsymbol{k}+\bs Q,\downarrow}
		\end{pmatrix}
		\end{equation}
		with
		\begin{align}
		\mathcal{H}^{ff}_{(1)}=
		\begin{pmatrix}
		\mathcal{Z}_+^{2} \xi^{f}_{\boldsymbol{k}} + \mathcal{Z}_-^{2} \xi^{f}_{\boldsymbol{k}+\bs Q} & 
		0 & 
		0 & 
		\mathcal{Z}_+ \mathcal{Z}_- (\xi^{f}_{\boldsymbol{k}+\bs Q} + \xi^{f}_{\boldsymbol{k}}) \\
		0 & 
		\mathcal{Z}_+^{2} \xi^{f}_{\boldsymbol{k}} + \mathcal{Z}_-^{2} \xi^{f}_{\boldsymbol{k}+\bs Q} & 
		\mathcal{Z}_+ \mathcal{Z}_- (\xi^{f}_{\boldsymbol{k}+\bs Q} + \xi^{f}_{\boldsymbol{k}}) & 
		0 \\
		0 & 
		\mathcal{Z}_+ \mathcal{Z}_- (\xi^{f}_{k} + \xi^{f}_{\boldsymbol{k}+\bs Q}) & 
		\mathcal{Z}_+^{2} \xi^{f}_{\boldsymbol{k}+\bs Q} + \mathcal{Z}_-^{2} \xi^{f}_{\boldsymbol{k}} & 
		0 \\
		\mathcal{Z}_+ \mathcal{Z}_- (\xi^{f}_{\boldsymbol{k}} + \xi^{f}_{\boldsymbol{k}+\bs Q}) & 
		0 & 
		0 & 
		\mathcal{Z}_+^{2} \xi^{f}_{\boldsymbol{k}+\bs Q} + \mathcal{Z}_-^{2} \xi^{f}_{\boldsymbol{k}}
		\end{pmatrix} \ .
		\label{eq:19lasgbv}
		\end{align}

		\textit{$\beta$-matrix coupling} ---
		We further investigate the fermionic $\beta$ Lagrange multiplier term
		\begin{align*}
		\beta
		\sum_{i}
		\bs f^\dagger_i
		\begin{pmatrix}
		0 & e^{ \textrm{i} \phi_{i}} \\
		e^{-\textrm{i} \phi_{i}} & 0
		\end{pmatrix}
		\bs f^\nodag_{i}
		&=
		\beta \frac{1}{N/2}
		\sum_{\boldsymbol{k},\boldsymbol{k}'}
		\sum_{i}
		e^{-\textrm{i}\boldsymbol{k} \boldsymbol{r}_{i}}
		e^{\textrm{i} \boldsymbol{k}' \boldsymbol{r}_{i}}
		\bs f^\dagger_{\bs k}
		\begin{pmatrix}
		0 & e^{\textrm{i} \bs Q \boldsymbol{r}_{i}} \\
		e^{-\textrm{i} \bs Q \boldsymbol{r}_{i}} & 0
		\end{pmatrix}
		\bs f^\nodag_{\bs k'}
		\\&=2
		\beta
		\sum_{\boldsymbol{k},\boldsymbol{k}'}
		\bs f^\dagger_{\bs k}
		\begin{pmatrix}
		0 & \delta_{\boldsymbol{k}',\boldsymbol{k}-\bs Q} \\
		\delta_{\boldsymbol{k}',\boldsymbol{k}+\bs Q} & 0
		\end{pmatrix}
		\bs f^\nodag_{\bs k'}
		\\
		&=2
		\beta
		\sum_{\boldsymbol{k}}
		(
		\boldsymbol{f}^{\dagger}_{\boldsymbol{k}}\, \tau^{+} \boldsymbol{f}_{\boldsymbol{k}-\bs Q}
		+
		\boldsymbol{f}^{\dagger}_{\boldsymbol{k}} \tau^{-} \boldsymbol{f}_{\boldsymbol{k}+\bs Q}
		) \ .
		\label{eq:35mteabr}
		\end{align*}
		After symmetrizing, and exploiting $\boldsymbol{k}-\bs Q \equiv \boldsymbol{k}+\bs Q$, we find
		\begin{equation}
		\sum_{i} 
		\bs f_i^\dagger
		\begin{pmatrix}
		\beta_0 & \beta e^{\textrm{i} \phi_{i}} \\
		\beta e^{-\textrm{i} \phi_{i}} & \beta_0
		\end{pmatrix}
		\bs f_i^\nodag
		= 
		\sum_{\boldsymbol{k}}
		\begin{pmatrix}
		f^{\dagger}_{\boldsymbol{k},\uparrow} &
		f^{\dagger}_{\boldsymbol{k},\downarrow} &
		f^{\dagger}_{\boldsymbol{k}+\bs Q,\uparrow} &
		f^{\dagger}_{\boldsymbol{k}+\bs Q,\downarrow}
		\end{pmatrix}
		\mathcal{H}^{ff}_{\beta}
		\begin{pmatrix}
		f_{\boldsymbol{k},\uparrow} \\
		f_{\boldsymbol{k},\downarrow} \\
		f_{\boldsymbol{k}+\bs Q,\uparrow} \\
		f_{\boldsymbol{k}+\bs Q,\downarrow}
		\end{pmatrix}
		\end{equation}
		with
		\begin{equation}
		\mathcal{H}^{ff}_{\beta}
		=
		\begin{pmatrix}
		\beta_0&0&0&\beta \\
		0&\beta_0&\beta&0 \\
		0&\beta&\beta_0&0 \\
		\beta&0&0&\beta_0
		\end{pmatrix}
		\ .
		\label{eq:35mteabr}
		\end{equation}
		\textit{Chemical potential term} ---
		Finally, we define
		\begin{equation}
		H_\mu=
		\sum_{\boldsymbol{k}}
		\begin{pmatrix}
		c^{\dagger}_{\boldsymbol{k},\uparrow} &
		c^{\dagger}_{\boldsymbol{k},\downarrow} &
		c^{\dagger}_{\boldsymbol{k}+\bs Q,\uparrow} &
		c^{\dagger}_{\boldsymbol{k}+\bs Q,\downarrow}
		\end{pmatrix}
		\mathcal{H}^{cc}_{\mu}
		\begin{pmatrix}
		c_{\boldsymbol{k},\uparrow} \\
		c_{\boldsymbol{k},\downarrow} \\
		c_{\boldsymbol{k}+\bs Q,\uparrow} \\
		c_{\boldsymbol{k}+\bs Q,\downarrow}
		\end{pmatrix}
		+
		\sum_{\boldsymbol{k}}
		\begin{pmatrix}
		f^{\dagger}_{\boldsymbol{k},\uparrow} &
		f^{\dagger}_{\boldsymbol{k},\downarrow} &
		f^{\dagger}_{\boldsymbol{k}+\bs Q,\uparrow} &
		f^{\dagger}_{\boldsymbol{k}+\bs Q,\downarrow}
		\end{pmatrix}
		\mathcal{H}^{ff}_{\mu}
		\begin{pmatrix}
		f_{\boldsymbol{k},\uparrow} \\
		f_{\boldsymbol{k},\downarrow} \\
		f_{\boldsymbol{k}+\bs Q,\uparrow} \\
		f_{\boldsymbol{k}+\bs Q,\downarrow}
		\end{pmatrix}
		\end{equation}
		with
		\begin{equation}
		\mathcal{H}^{cc}_{\mu}=
		\begin{pmatrix}
		-\mu_0&0&0&0\\
		0&-\mu_0&0&0\\
		0&0&-\mu_0&0\\
		0&0&0&-\mu_0
		\end{pmatrix}
		\ , \quad \mathcal{H}^{ff}_{\mu}=
		\begin{pmatrix}
		-\mu_0+\epsilon_f&0&0&0\\
		0&-\mu_0+\epsilon_f&0&0\\
		0&0&-\mu_0+\epsilon_f&0\\
		0&0&0&-\mu_0+\epsilon_f
		\end{pmatrix} \ .
		\end{equation}
		\textit{Result} ---
		As a result, the full form of the mean-field Hamiltonian is expressed as
		\begin{equation}
		\tilde{H}= \sum_{\bs k} \bs \Phi_{\bs k}^\dagger \mathcal{H}_{\bs k, \bs Q}^{8 \times 8} \bs \Phi_{\bs k}^\nodag
		\  \quad \text{with}  \quad
		\mathcal{H}_{\bs k, \bs Q}^{8 \times 8}
		=
		\begin{pmatrix}
		\mathcal{H}^{cc} & \mathcal{H}^{cf} \\
		\mathcal{H}^{fc} & \mathcal{H}^{ff}
		\end{pmatrix}
		\ , \quad \mathcal{H}^{cc} = \mathcal{H}^{cc}_{t} + \mathcal{H}^{cc}_{\mu} \ , \quad
		\mathcal{H}^{ff} = \mathcal{H}^{ff}_{t} + \mathcal{H}^{ff}_{\beta}+ \mathcal{H}^{ff}_{\mu} \ .
		\end{equation}
		in the basis of Eq.~\eqref{eq:5uktvd}.
		To calculate topological invariants one has to rotate the basis of the hopping Hamiltonian
		\begin{equation}
		 \mathcal{H}_{B,\bs k, \bs Q}^{8 \times 8}= U^\dagger_{B,\bs k}\mathcal{H}_{\bs k, \bs Q}^{8 \times 8}U^\nodag_{B,\bs k} \label{eq:BlochH},	
\end{equation}		
to get the Bloch form, which obeys $\mathcal{H}_{B,\bs k, \bs Q}^{8 \times 8}=\mathcal{H}_{B,\bs k+ \bs Q, \bs Q}^{8 \times 8}$, e.g. 
		\begin{align*}
		\boldsymbol{c}_{1,\boldsymbol k}^\dagger =\phantom{-} \frac{1}{\sqrt{2}} e^{\text{i}k_x}	\boldsymbol{c}_{\boldsymbol k}^\dagger + 	 \frac{1}{\sqrt{2}} \boldsymbol{c}_{\boldsymbol k+\boldsymbol Q}^\dagger \  , \\
		\boldsymbol{c}_{2,\boldsymbol k}^\dagger = -\frac{1}{\sqrt{2}} e^{\text{i}k_x}	\boldsymbol{c}_{\boldsymbol k}^\dagger + 	 \frac{1}{\sqrt{2}} \boldsymbol{c}_{\boldsymbol k+\boldsymbol Q}^\dagger \ , \\
		\boldsymbol{f}_{1,\boldsymbol k}^\dagger =\phantom{-} \frac{1}{\sqrt{2}} e^{\text{i}k_x}	\boldsymbol{f}_{\boldsymbol k}^\dagger + 	 \frac{1}{\sqrt{2}} \boldsymbol{f}_{\boldsymbol k+\boldsymbol Q}^\dagger \ , \\
		\boldsymbol{f}_{2,\boldsymbol k}^\dagger = -\frac{1}{\sqrt{2}} e^{\text{i}k_x}	\boldsymbol{f}_{\boldsymbol k}^\dagger + 	 \frac{1}{\sqrt{2}} \boldsymbol{f}_{\boldsymbol k+\boldsymbol Q}^\dagger \ ,
		\end{align*}
		which can be expressed as
		\begin{equation}
		 \bs \Phi_{B,\bs k}^\dagger = U^\dagger_{B,\bs k}  \bs \Phi_{\bs k}^\dagger,
		\end{equation}
		where $\bs \Phi_{B,\bs k}^\dagger$ is the rotated basis. We hence refer to to this rotated hopping  
	The inversion $\mathcal{I}$ operator is given by
		\begin{equation}
		 \mathcal{I} = U^\dagger_{B,-\bs k} 		
		\begin{pmatrix}
		1 & 0 & 0 & 0 & 0 & 0 & 0 & 0 \\
		0 & 1 & 0 & 0 & 0 & 0 & 0 & 0 \\
		0 & 0 & 1 & 0 & 0 & 0 & 0 & 0 \\
		0 & 0 & 0 & 1 & 0 & 0 & 0 & 0 \\
		0 & 0 & 0 & 0 & -1 & 0 & 0 & 0 \\
		0 & 0 & 0 & 0 & 0 & -1 & 0 & 0 \\
		0 & 0 & 0 & 0 & 0 & 0 & -1 & 0 \\
		0 & 0 & 0 & 0 & 0 & 0 & 0 & -1 \\
		\end{pmatrix}
		 U^\nodag_{B,\bs k} 	
		\end{equation}
		and the inversion eigenvalues of the n-th Kramer's pair at time invariant momentum $\Gamma_{j}$ can be obtained by calculating
		\begin{equation}
		 \xi[n,\Gamma_{j}]=\braket{\pm\sigma, n, -\bs k|\mathcal{I}|\pm\sigma, n, \bs k},
		\end{equation}
		where $\ket{\pm\sigma, n, \bs k}$ represents the two eigenvectors for the nth Kramer's pair of the hopping matrix $ \mathcal{H}_{B,\bs k, \bs Q}^{8 \times 8}$, defined in Eq.~\eqref{eq:BlochH}, with wave vector $\bs k$ and pseudo spin $\sigma$.

		For $\bs Q = 0$, we can either get a paramagnetic $p = 0$ ($\beta = 0$) or ferromagnetic $p \neq 0$ ($\beta \neq 0$) solutions. In this case, the system has the generic BZ with $k_\alpha \in \left\{-\pi,\pi\right\}$. Consequently we can write the Hamiltonian
		as a $4\times 4 $ hopping matrix with the basis 
		\begin{equation}
		\bs \varphi_{\bs k}^\dagger =
		\begin{pmatrix}
		c_{\boldsymbol{k},\uparrow}^{\dagger} & 
		c_{\boldsymbol{k},\downarrow}^{\dagger} &
		f_{\boldsymbol{k},\uparrow}^{\dagger} &
		f_{\boldsymbol{k},\downarrow}^{\dagger} 
		\end{pmatrix} \ .
		\end{equation}
		
		We find 
				\begin{equation}
			\tilde{H}= \sum_{\bs k} \bs \varphi_{\bs k}^\dagger \mathcal{H}_{\bs k, 0}^{4 \times 4} \bs \varphi_{\bs k}^\nodag
			\  \quad \text{with}  \quad
			\mathcal{H}_{\bs k,0}^{4 \times 4}
			=
			\begin{pmatrix}
				\mathcal{H}^{cc}_0 & \mathcal{H}^{cf}_0 \\
				\mathcal{H}^{fc}_0 & \mathcal{H}^{ff}_0
			\end{pmatrix}
		\end{equation}
	and
		\begin{equation}
		\begin{split}
			\mathcal{H}^{cc}_0
		&=
		\begin{pmatrix}
		\xi^{c}_{\boldsymbol{k}} - \mu& 0 \\
		0 & \xi^{c}_{\boldsymbol{k}} -\mu
		\end{pmatrix} \ , \\
		\mathcal{H}^{cf}_0
		&=
		\left( \mathcal{Z}_+ \tau^\alpha s_{\alpha,\bs k} +\mathcal{Z}_- \tau^\alpha \tau^1 s_{\alpha,\bs k} \right) , 
		\\
	\mathcal{H}^{ff}_0 &=
		\begin{pmatrix}
		\mathcal{Z}_+^2 \xi^{f}_{\boldsymbol{k}} + \mathcal{Z}_-^2 \xi^{f}_{\boldsymbol{k}} - \mu + \beta_0 + \epsilon_f 
		& 2 \mathcal{Z}_+\mathcal{Z}_- \xi^{f}_{\boldsymbol{k}} + \beta \\
		2 \mathcal{Z}_+\mathcal{Z}_- \xi^{f}_{\boldsymbol{k}} + \beta 
		& \mathcal{Z}_+^2 \xi^{f}_{\boldsymbol{k}} + \mathcal{Z}_-^2 \xi^{f}_{\boldsymbol{k}} - \mu + \beta_0 + \epsilon_f 
		\end{pmatrix} \ .
		\end{split}
		\end{equation}
	The hopping Hamiltonian $\mathcal{H}_{\bs k,0}^{4 \times 4}$ is a Bloch form and its respective inversion operator $\mathcal{I}$ reads
		\begin{equation}
		\mathcal{I} =
		\begin{pmatrix}
		1 & 0 & 0 & 0  \\
		0 & 1 & 0 & 0  \\
		0 & 0 & -1 & 0  \\
		0 & 0 & 0 & -1  \\
		\end{pmatrix}
		\end{equation}
			
		\subsection{Rotation of the spin orientation in the magnetic MF ansatz}
		With the mean field ansatz, we assumed a magnetic spiral, that rotates in $xy$-plane, i.e.,
		\begin{equation}
		\boldsymbol{p}_{i}
		= (p\cos \phi_{i}, p\sin \phi_{i}, 0)
		= (p\cos \bs Q \boldsymbol{r}_{i}, p\sin \bs Q \boldsymbol{r}_{i},0).
		\end{equation}
	In the following, we show that the direction of the spin spiral, does not change the mean field solution, i.e. the above ansatz is without loss of generality.
	We do so, by applying a general rotation of the the Pauli matrices $\bs \tau$
		\begin{align}
		\begin{split}
		\begin{pmatrix}
		\tilde{f}_{i,\uparrow}\\ \tilde{f}_{i, \downarrow} 
		\end{pmatrix} 
		&\rightarrow
		U^\dagger(\hat{\boldsymbol{n}},\theta)
		\begin{pmatrix}
		\mathcal{Z}_+ & \mathcal{Z}_-\, e^{-\textrm{i} \phi_i} \\
		\mathcal{Z}_-\, e^{\textrm{i} \phi_i} & \mathcal{Z}_+
		\end{pmatrix}
		U(\hat{\boldsymbol{n}},\theta)
		\begin{pmatrix}
		f_{i,\uparrow}\\ f_{i, \downarrow}
		\end{pmatrix}  \ ,
		\\
		\begin{pmatrix}
		0 & \beta e^{-\textrm{i} \phi_i} \\
		\beta e^{\textrm{i} \phi_i} & 0
		\end{pmatrix}
		&\rightarrow
		U^\dagger(\hat{\boldsymbol{n}},\theta)
		\begin{pmatrix}
		0 & \beta e^{-\textrm{i} \phi_i} \\
		\beta e^{\textrm{i} \phi_i} & 0
		\end{pmatrix}
		U(\hat{\boldsymbol{n}},\theta) \quad \text{with} \quad 
		U(\hat{\boldsymbol{n}},\theta) 
		:= e^{-\textrm{i} \frac{\theta}{2} (n^{1} \sigma^{1}+n^{2} \sigma^{2}+n^{3} \sigma^{3})} \ ,
		\end{split}
		\end{align}
	where $\theta$ denotes the rotation angle around the axis $\hat{\boldsymbol{n}} := (n^{1},n^{2},n^{3})$.
		Within the basis of Eq.~\eqref{eq:5uktvd} the rotated Hamiltonian is given by,
		\begin{equation}
		\tilde{H}= \sum_{\bs k} \bs \Phi_{\bs k}^\dagger \mathcal{H}_{\bs k, \bs Q}^{8 \times 8} \bs \Phi_{\bs k}^\nodag
		\  \quad \text{with}  \quad
		\mathcal{H}_{\bs k, \bs Q}^{8 \times 8}
		=
		\begin{pmatrix}
		\mathcal{H}^{cc} & \mathcal{H}^{cf} \\
		\mathcal{H}^{fc} & \mathcal{H}^{ff}
		\end{pmatrix}
		\end{equation}
		where
		\begin{equation}
		\begin{split}
		\mathcal{H}^{cc}
		& \rightarrow
		\begin{pmatrix}
		(\xi^{c}_{\boldsymbol{k}} -\mu)\tau^0 & 0 \\
		0 & (\xi^{c}_{\boldsymbol{k}+\bs Q}-\mu)\tau^0 \\
		\end{pmatrix} \ , \\
		\mathcal{H}^{fc} & \rightarrow -V \sum_{\alpha}
		\begin{pmatrix}
		\mathcal{Z}_+ s_{\alpha,\bs k} \tau^\alpha & 
		\mathcal{Z}_- s_{\alpha,\bs k+\bs Q}
		U^\dagger(\hat{\boldsymbol{n}},\theta)\, \tau^{1}\, U(\hat{\boldsymbol{n}},\theta) \tau^\alpha \\
		\mathcal{Z}_- s_{\alpha,\bs k}
		U^\dagger(\hat{\boldsymbol{n}},\theta)\, \tau^{1}\, U(\hat{\boldsymbol{n}},\theta) \tau^\alpha & \mathcal{Z}_+s_{\alpha,\bs k+\bs Q} \tau^\alpha \\
		\end{pmatrix} \ , \\
		\mathcal{H}^{ff} & \rightarrow 
		\begin{pmatrix}
		\big(\mathcal{Z}_+^{2} \xi^{f}_{\boldsymbol{k}} + \mathcal{Z}_-^{2} \xi^{f}_{\boldsymbol{k}+\bs Q} +\beta_{0} -\mu \big)\, \tau^{0} &
		\big(\mathcal{Z}_+ \mathcal{Z}_- (\xi^{f}_{\boldsymbol{k}+\bs Q} + \xi^{f}_{\boldsymbol{k}}) +\beta \big)\, U^\dagger(\hat{\boldsymbol{n}},\theta)\, \tau^{1}\, U(\hat{\boldsymbol{n}},\theta)\\
		\big(\mathcal{Z}_+ \mathcal{Z}_- (\xi^{f}_{\boldsymbol{k}} + \xi^{f}_{\boldsymbol{k}+\bs Q}) +\beta \big)\, U^\dagger(\hat{\boldsymbol{n}},\theta)\, \tau^{1}\, U(\hat{\boldsymbol{n}},\theta)& 
		\big(\mathcal{Z}_+^{2} \xi^{f}_{\boldsymbol{k}+\bs Q} + \mathcal{Z}_-^{2} \xi^{f}_{\boldsymbol{k}} +\beta_{0} -\mu \big)\, \tau^{0}
		\end{pmatrix} \ .
		\end{split}
		\end{equation}
		We identify, that rotating the magnetization plane of the mean field ansatz is equivalent to rotating the direction of spin-orbit coupling by setting
		$\tau^\alpha \rightarrow U^\dagger(\hat{\boldsymbol{n}},\theta)\, \tau^{\alpha}\, U(\hat{\boldsymbol{n}},\theta)$
		in Eq.~\ref{app:Hamiltonian}. It can shown by explicit calculation, that the eigenvalues of $\mathcal{H}_{\bs k, \bs Q}^{8 \times 8} $ are independent of $U^\dagger(\hat{\boldsymbol{n}},\theta)$. The physical reason is, that in the presence of a time reversal and inversion symmetry, such a rotation can not change
		the eigenspectrum, since it reduces to rotating in the degenerate subspace of the Hamiltonian
		\begin{equation}
		U(\hat{\boldsymbol{n}},\theta) \ket{\sigma, n, \bs k} = u_1(\boldsymbol{n},\theta) \ket{\sigma, n, \bs k} + u_2(\boldsymbol{n},\theta)\ket{-\sigma, n, \bs k}
		\quad \text{with} \quad \mathcal{H}_{\bs k, \bs Q}^{8 \times 8}\ket{\pm\sigma, n, \bs k} = \epsilon^\nu_{\bs k,\bs Q} \ket{\pm\sigma, n, \bs k},
		\end{equation}
		where $\ket{\pm\sigma, n, \bs k}$ represents the two eigenvectors for the nth Kramer's pair with the degenerate eigenvalue $\epsilon^\nu_{\bs k,\bs Q}$, wave vector $\bs k$ and pseudo spin $\sigma$.

		\subsection{Mean field with arbitrary \texorpdfstring{$\bs Q$}{}-vector}
		We present effective fermionic Hamiltonian matrix $H^{\textrm{hop};N^{0}}_{\boldsymbol{k}}$ with more general $\bs Q$ such that $N^{0} \bs Q \equiv \boldsymbol{G}$ ($N^0 \in \mathbb{N}$).
		For $N^{0} > 2$, due to the reduced translational symmetries, the matrix is of a bigger size of $4N^{0}$ by $4N^{0}$.
		Note that, for incommensurate ordering vectors with irrational or random period, the calculation below does not apply.
		
		In the previous calculations, without $2 \bs Q \equiv 0 \textrm{ mod } \boldsymbol{G}$, one finds couplings of forms
		\begin{subequations}
			\begin{align}
			\boldsymbol c^\dagger_{\boldsymbol k} \mathcal{H}^{cc}_{\boldsymbol k} \boldsymbol c_{\boldsymbol k} 
			&\quad \text{with} \quad
			\mathcal{H}^{cc}_{\boldsymbol k} = \xi^{c}_{\boldsymbol{k}}\, \tau^{0} \ ,
			\\
			\boldsymbol f^\dagger_{\boldsymbol k} \mathcal{H}^{fc,1}_{\boldsymbol k} \boldsymbol c_{\boldsymbol k}
			&\quad \text{with} \quad
			\mathcal{H}^{fc,1}_{\boldsymbol k}= -2V\mathcal{Z}_+\sin(\boldsymbol{k}\hat{\boldsymbol{r}}_{\alpha})\, \tau^\alpha \ ,
			\\
			\boldsymbol f^\dagger_{\boldsymbol k} \mathcal{H}^{ff,1}_{\boldsymbol k} \boldsymbol f_{\boldsymbol k} 
			&\quad \text{with} \quad
			\mathcal{H}^{ff,1}_{\boldsymbol k}=
			\mathcal{Z}_+^{2} \xi^{f}_{\boldsymbol{k}}\, \tau^{0}
			+
			\mathcal{Z}_-^{2}
			\begin{pmatrix}
			\xi^{f}_{\boldsymbol{k}+\bs Q} & 0 \\
			0 & \xi^{f}_{\boldsymbol{k}-\bs Q}
			\end{pmatrix}
			\ ,
			\\
			\boldsymbol f^\dagger_{\boldsymbol k+ \boldsymbol Q} \mathcal{H}^{fc,2}_{\boldsymbol k} \boldsymbol c_{\boldsymbol k}
			&\quad \text{with} \quad
			\mathcal{H}^{fc,2}_{\boldsymbol k}= -2V\mathcal{Z}_- \sin (\boldsymbol{k}\hat{\boldsymbol{r}}_{\alpha})\, \tau^{-}\tau^\alpha \ ,
			\\
			\boldsymbol f^\dagger_{\boldsymbol k+\boldsymbol Q} \mathcal{H}^{ff,2}_{\boldsymbol k} \boldsymbol f_{\boldsymbol k}
			&\quad \text{with} \quad
			\mathcal{H}^{ff,2}_{\boldsymbol k}= \mathcal{Z}_+\mathcal{Z}_-(\xi^f_{\boldsymbol k}+\xi^f_{\boldsymbol k+\boldsymbol Q})\,
			\tau^{-}
			+\beta \tau^{+} \ ,
			\\
			\boldsymbol f^\dagger_{\boldsymbol k} \mathcal{H}^{fc,3}_{\boldsymbol k} \boldsymbol c_{\boldsymbol k+\bs Q}
			&\quad \text{with} \quad
			\mathcal{H}^{fc,3}_{\boldsymbol k}= -2V\mathcal{Z}_- \sin ((\boldsymbol{k}+\bs Q) \hat{\boldsymbol{r}}_{\alpha})\, \tau^{+}\tau^\alpha \ .
			\end{align}
			\label{eq:30ntemnym}
		\end{subequations}
		Similarly, the matrices
		\begin{equation*}
		\mathcal{H}^{cf,1}_{\boldsymbol{k}} = (\mathcal{H}^{fc,1})^{\dagger} \ , \quad
		\mathcal{H}^{cf,2}_{\boldsymbol{k}} = (\mathcal{H}^{fc,2})^{\dagger} \ , \quad
		\mathcal{H}^{cf,3}_{\boldsymbol{k}} = (\mathcal{H}^{fc,3})^{\dagger} \ , \quad
		\mathcal{H}^{ff',2}_{\boldsymbol{k}}= (\mathcal{H}^{ff,2})^{\dagger}
		\end{equation*}
		are defined to express the Hermitian conjugate of relevant ones in Eq.~\eqref{eq:30ntemnym}.
		We now plug in our ansatz $N^{0} \bs Q \equiv 0 \textrm{ mod } \boldsymbol{G}$.
		Using the basis
		\begin{equation*}
		\begin{pmatrix}
		\boldsymbol c^\dagger_{\boldsymbol k} &  \boldsymbol f^\dagger_{\boldsymbol k} &  \boldsymbol c^\dagger_{\boldsymbol k + \boldsymbol Q} &  \boldsymbol f^\dagger_{\boldsymbol k + \boldsymbol Q}
		& \boldsymbol c^\dagger_{\boldsymbol k + 2\boldsymbol Q} &  \boldsymbol f^\dagger_{\boldsymbol k + 2\boldsymbol Q} & ... &  \boldsymbol c^\dagger_{\boldsymbol k + (N-1)\boldsymbol Q} &  \boldsymbol f^\dagger_{\boldsymbol k + (N-1)\boldsymbol Q}
		\end{pmatrix}
		\ ,
		\end{equation*}
		the Hamilton matrix $H^{\textrm{hop};N^{0}}_{\boldsymbol{k}}$ is expressed by
		\begin{equation}
		\frac{1}{N^{0}}
		\left( \begin{array}{ccccccccccc}
		\mathcal{H}^{cc}_{\boldsymbol k} & \mathcal{H}^{cf,1}_{\boldsymbol k} & 0 &  \mathcal{H}^{cf,2}_{\boldsymbol k} & 0 & 0 & \cdots & 0 & 0 & 0 & \mathcal{H}^{cf,3}_{\boldsymbol k-\bs Q}\\
		\mathcal{H}^{fc,1}_{\boldsymbol k} & \mathcal{H}^{ff,1}_{\boldsymbol k}  & \mathcal{H}^{fc,3}_{\boldsymbol k} & \mathcal{H}^{ff',2}_{\boldsymbol k} & 0 & 0 & \cdots & 0 & 0 & H^{fc,2}_{\boldsymbol k - \boldsymbol Q} & \mathcal{H}^{ff,2}_{\boldsymbol k - \boldsymbol Q} \\
		0 & \mathcal{H}^{cf,3}_{\boldsymbol k} & \mathcal{H}^{cc}_{\boldsymbol k + \boldsymbol Q} & \mathcal{H}^{cf,1}_{\boldsymbol k + \boldsymbol Q} &  0 & \mathcal{H}^{cf,2}_{\boldsymbol k + \boldsymbol Q} & \cdots & 0 & 0 & 0 & 0 \\
		\mathcal{H}^{fc,2}_{\boldsymbol k} & \mathcal{H}^{ff,2}_{\boldsymbol k} & \mathcal{H}^{fc,1}_{\boldsymbol k + \boldsymbol Q} & \mathcal{H}^{ff,1}_{\boldsymbol k+ \boldsymbol Q}  & \mathcal{H}^{fc,3}_{\boldsymbol k + \boldsymbol Q} & \mathcal{H}^{ff',2}_{\boldsymbol k + \boldsymbol Q} & \cdots & 0 & 0 & 0 & 0 \\
		0 & 0 & 0 & \mathcal{H}^{cf,3}_{\boldsymbol k +\boldsymbol Q} & \mathcal{H}^{cc}_{\boldsymbol k + 2\boldsymbol Q} & \mathcal{H}^{cf,1}_{\boldsymbol k + 2\boldsymbol Q} & \cdots & 0 & 0  & 0 & 0 \\
		0 & 0 & \mathcal{H}^{fc,2}_{\boldsymbol{k}+\bs Q} & \mathcal{H}^{ff,2}_{\boldsymbol{k}+\bs Q} & \mathcal{H}^{fc,1}_{\boldsymbol{k}+2\bs Q} & \mathcal{H}^{ff,1}_{\boldsymbol{k}+2\bs Q} & \cdots & 0 & 0 & 0 & 0 \\
		\vdots & \vdots & \vdots & \vdots & \vdots & \vdots & \ddots & \vdots & \vdots & \vdots & \vdots \\
		0&0&0&0&0&0& \cdots &\mathcal{H}^{cc}_{\boldsymbol{k}+(N^0-2)\bs Q} & \mathcal{H}^{cf,1}_{\boldsymbol{k}+(N^0-2)\bs Q} & 0 & \mathcal{H}^{cf,2}_{\boldsymbol{k}+(N^0-2)\bs Q}\\
		0&0&0&0&0&0& \cdots & \mathcal{H}^{fc,1}_{\boldsymbol{k}+(N^0-2)\bs Q} & \mathcal{H}^{ff,1}_{\boldsymbol{k}+(N^0-2)\bs Q} & \mathcal{H}^{fc,3}_{\boldsymbol{k}+(N^0-2)\bs Q} & H^{ff',2}_{\boldsymbol{k}+(N^0-2)\bs Q} \\
		0 &  \mathcal{H}^{cf,2}_{\boldsymbol k - \boldsymbol Q} & 0 & 0 & 0 & 0 & \cdots & 0 &  \mathcal{H}^{cf,3}_{\boldsymbol k +(N^0-2)\boldsymbol Q} & \mathcal{H}^{cc}_{\boldsymbol k + (N^0-1)\boldsymbol Q} & \mathcal{H}^{cf,1}_{\boldsymbol k + (N^0-1)\boldsymbol Q} \\
		\mathcal{H}^{fc,3}_{\boldsymbol k-\bs Q} & \mathcal{H}^{ff',2}_{\boldsymbol k - \boldsymbol Q} & 0 & 0 & 0 & 0 & \cdots & \mathcal{H}^{fc,2}_{\boldsymbol k +(N^0-2)\boldsymbol Q} & \mathcal{H}^{ff,2}_{\boldsymbol k +(N^0-2)\boldsymbol Q} & \mathcal{H}^{fc,1}_{\boldsymbol k + (N^0-1)\boldsymbol Q} & \mathcal{H}^{ff,1}_{\boldsymbol k+ (N^0-1)\boldsymbol Q}
		\end{array} \right) \ .
		\end{equation}

\subsection{Numerical probe of excitonic states}
We investigate the existence of excitonic states by computing the imaginary part of the spin susceptibility as a function of the frequency of the external field $\omega$.
We adopt the analytical form in the scheme of slave-bosonic representation, which is derived in \cite{PAM_Wuerzburg}.
To see the instability relevant to the spontaneous phase transition to AFM, we implement the calculation in the paramagnetic phase.
As Fig.~\ref{fig:7nhehhrjb} illustrates, we find only one peak, which appears below the nesting frequency with $\boldsymbol{Q}=(\pi,\pi,\pi)$ and drives the AFM phase with higher interaction parameter $U$.
Therefore, we conclude that either (i) the system does not have excitonic states, or (ii) slave-boson approach is not capable of finding them.

\begin{figure}[h]
\centering
\includegraphics[width=0.46\textwidth]{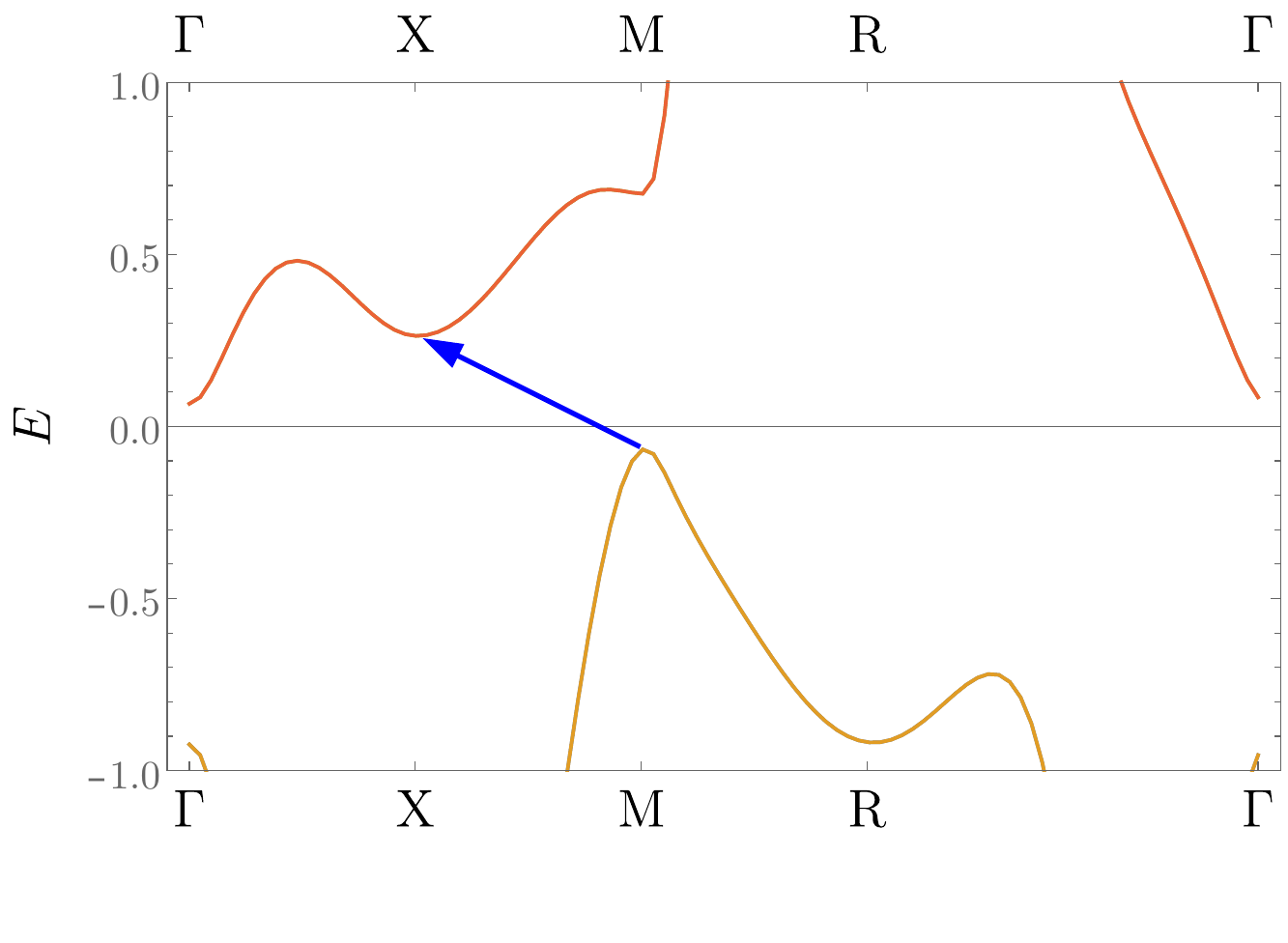}
$\quad$
\includegraphics[width=0.46\textwidth]{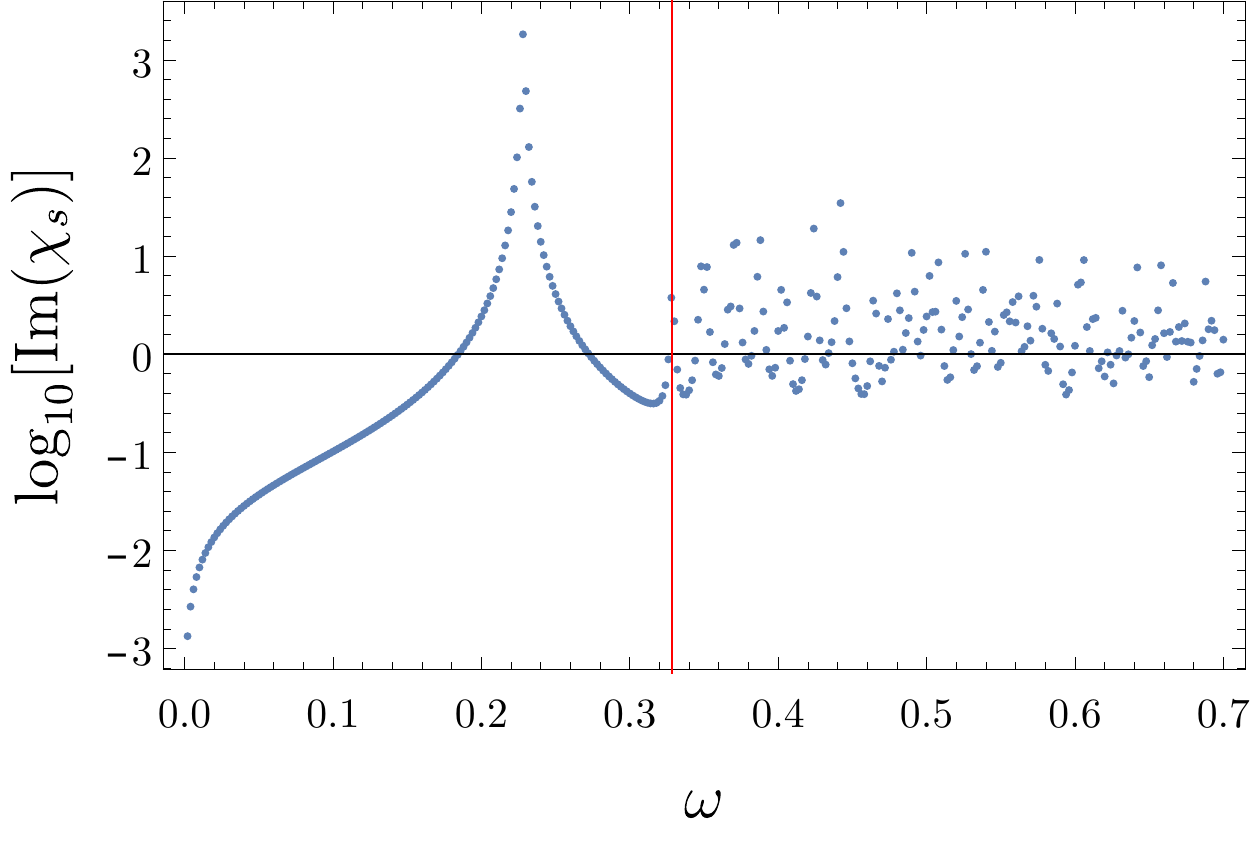}
\caption{
\textit{Left}: Effective band structure at $(\epsilon_f,U)=(-2,4)$ in the paramagnetic phase.
Nesting with momentum $\boldsymbol{Q}=(\pi,\pi,\pi)$ occurs at $\omega \approx 0.32$ (blue arrow) or higher.
\textit{Right}: Spin susceptibility with $\boldsymbol{Q}=(\pi,\pi,\pi)$ as a function of $\omega$ at $(\epsilon_f,U)=(-2,4)$.
A sharp divergence at around $\omega \approx 0.22$ triggers AFM phase at higher $U$.
Fluctuations are observed on the right side of the vertical red line, where the nesting condition is fulfilled.
}
\label{fig:7nhehhrjb}
\end{figure}

\end{widetext}

\end{document}